\documentclass[fleqn,usenatbib,useAMS]{mnras}

\usepackage[T1]{fontenc}
\usepackage{ae,aecompl}
\usepackage{graphicx}	
\usepackage{amsmath}	
\usepackage{amssymb}	
\usepackage{enumitem}
\usepackage{rotating}
\usepackage{pdflscape}
\usepackage{newtxtext,newtxmath}

\newcommand{\gc}{$\gamma$\,Cas}

\title[Be stars in eROSITA]{SRG/eROSITA survey of Be stars\thanks{Based on spectra obtained with eROSITA }}

\author[Naz\'e \& Robrade]{Ya\"el~Naz\'e$^1$\thanks{F.R.S.-FNRS Senior Research Associate, email: ynaze@uliege.be}, Jan Robrade$^2$ 
\\
$^1$ Groupe d'Astrophysique des Hautes Energies, STAR, Universit\'e de Li\`ege, Quartier Agora (B5c, Institut d'Astrophysique et de G\'eophysique), \\
All\'ee du 6 Ao\^ut 19c, B-4000 Sart Tilman, Li\`ege, Belgium\\
$^2$ Hamburger Sternwarte, University of Hamburg, Gojenbergsweg 112, D-21029 Hamburg, Germany
}

\pubyear{2021}

\begin{document}
\label{firstpage}
\pagerange{\pageref{firstpage}--\pageref{lastpage}}
\maketitle

\begin{abstract}
Massive stars are known X-ray emitters and those belonging to the Be category are no exception. One type of X-ray emission even appears specific to that category, the \gc\ phenomenon. Its actual incidence has been particularly difficult to assess. Thanks to four semesters of sky survey data taken by SRG (Spectrum Roentgen Gamma)/eROSITA, we revisit the question of the X-ray properties of Be stars. Amongst a large catalog of Be stars, eROSITA achieved 170 detections (20\% of sample), mostly corresponding to the earliest spectral types and/or close objects. While X-ray luminosities show an uninterrupted increasing trend with the X-ray-to-bolometric luminosity ratios, the X-ray hardness was split between a large group of soft (and fainter on average) sources and a smaller group of hard (and brighter on average) sources. The latter category gathers at least 34 sources, nearly all displaying early spectral types. Only a third of them were known before to display such X-ray properties. The actual incidence of hard and bright X-rays amongst early-type Be stars within 100--1000\,pc appears to be $\sim$12\%, which is far from negligible. At the other extreme, no bright supersoft X-ray emission seem to be associated to any of our targets.
\end{abstract}


\section{Introduction}

The Be star category gathers early-type stars, at or close to the main sequence, that have shown, at least at some point, emission lines (such as in the Balmer series of Hydrogen or in Fe\,{\sc ii}) - for a review, see \citet{riv13}. These emissions were interpreted as evidence for a {\it decretion} disk surrounding the star, whose presence was confirmed by interferometric measurements \citep[e.g.][]{ste12,mou15}. Such detailed studies also revealed that the disks exhibit Keplerian motions.

The origin of these disks remains debated, but ejection of photospheric material through non-radial pulsations is often advocated \citep{hua09}. Such an equatorial ejection is facilitated by the fast rotation of the Be stars \citep{zor16,zor17}. This rotation could arise in two different ways. First, since stars are born with a range of rotational velocities, Be stars could simply be stars with a high initial stellar rotation rate, possibly spinned-up during main sequence evolution \citep{eks08}. Second, the fast rotation could be acquired after mass transfer in a close binary system. The Be star would then be the mass gainer, i.e. the initially less massive star of the system, and the companion a stripped star or a compact object (WD, NS, BH). Recent observational evidence suggests a high contribution of the binary channel \citep[e.g.][]{bou18,kle19}. Theoretical simulations have then suggested that Be stars with NS and BH companions would represent a tiny fraction of the Be population, while WD and stripped-star companions would be equally common \citep{sha14}. Detection of Be+WD systems remain elusive in the low-energy domain, but stripped-star companions have been detected in several cases thanks to their UV emission \citep{wan21}. However, observations at high energy can greatly help clarifying the situation.

Indeed, in the X-ray domain, classical Be stars were found to display various properties. When paired with NS, Be stars will form high-mass X-ray binaries (HMXBs) hence display very bright and hard X-rays (e.g. \citealt{wal15}). In addition, a few sources displaying supersoft X-ray emissions were identified as Be stars paired with WDs \citep{kah06,stu12,coe20,ken21}. Be+BH pairs are more elusive, with proposed candidates often discarded afterwards (see e.g. the case of MWC\,656 in \citealt{mun14} vs. \citealt{riv23}). Finally, the subcategory of \gc\ stars could be identified amongst Be stars (for a review, see \citealt{smi16}). Such objects display a thermal spectrum at high energies, but with unusually high plasma temperatures for massive stars ($kT>5$\,keV vs. $kT<2$\,keV) and X-ray luminosities intermediate between "normal" massive stars and HMXBs ($\log[L_{\rm X}/L_{\rm BOL}]$ between --6.2 and --4). Apart from their peculiar X-ray emissions, \gc\ stars are found to be quite typical Be stars in terms of photometric variations \citep{naz20tess} or multiplicity properties \citep{naz22} so that X-rays are required to identify these objects. Since most \gc\ discoveries were made serendipitously, their exact incidence rate remains to be determined, as well as their relationship with the other Be stars in the X-ray domain (clear-cut separation of the two categories or a continuum of X-ray properties?). 

A first dedicated survey of 84 Oe and Be stars not known to be HMXBs was performed by \citet{naz18} using archival XMM and Chandra data. They found many new cases of \gc\ stars; currently, about two dozens are known. They made the first steps towards a better understanding of the X-ray properties of Be stars, but the nature of the study (archival data) implies that biases may exist. In contrast, ROSAT did an overall survey which could have provided a clean sample, but it was focused on the soft X-ray range, while many phenomena in massive stars, notably the \gc\ phenomenon, mostly affect higher energies. Thanks to its sensitive survey of the X-ray sky up to the hard X-ray range, eROSITA can fill this gap and provide, for the first time, clear statistics regarding the high-energy properties of Be stars.

Section 2 presents the sample, Section 3 the data used for the study, while Section 4 discusses the results and Section 5 summarizes them.

\section{The sample}

To get a (quasi) complete list of Be stars, we rely on the Be Star Spectra (BeSS) catalog\footnote{http://basebe.obspm.fr/} \citep{nei11}. Since the exposures of the eROSITA survey are shallow, we decided to focus on the Milky Way, and excluded the Magellanic Cloud stars. We also excluded stars not listed as "classical Be stars" in that catalog, or classified in Simbad as HMXBs\footnote{Excluded cases for that reason: RX J0209.6--7427, CAS GAM-1, V831 Cas, V615 Cas, BQ Cam, HD 249179, V725 Tau, V420 Aur, RX J065817.7--071228, HD 259440, V572 Pup, 2E 1048.1--5937, HD 100199, 2E 1118.7--6138, 4U 1416-62, CPD--63 2495, V850 Cen, [BM83] X1553-542, RX J1739.4--2942,GSC 03588--00834,  HD 153295, GRO J2058+42, V2246 Cyg, 4U 2238+60, BD+53 2790, V2175 Cyg, V490 Cep, HD109857, V801 Cen}, young objects\footnote{Excluded cases: GSC 02342--00359, V743 Mon, Cl* NGC 2244 JOHN 33, GU CMa, 17 Sex, Cl* Trumpler 16 MJ 632, Cl* NGC 6530 ZCW 228, Cl* NGC 6530 ZCW 221, Cl* NGC 6530 ZCW 175, EM* LkHA 133, BD+41 3731, V2018 Cyg, V1493 Cyg, V385 Cep, V374 Cep, EM* LkHA 201, Cl* NGC 6530 ZCW 254, Cl* NGC 6611 ESL 22, BD--13 4936} or massive stars of other nature\footnote{Excluded cases: the LBVs HR Car and AG Car, the Mira SS73 167, and the WR CSI-62-12087 and SS73 123}.

With Vizier\footnote{https://vizier.cds.unistra.fr/}, we cross-correlated our BeSS list with {\it GAIA} DR3 catalogs, and kept only stars for which the counterpart had a secure distance, i.e. the ratio between the parallax and its error $R_{plx}>5$ and a re-normalized unit weight error $RUWE<1.4$, to avoid problems. The final distance values were taken from the median of the photogeometric distance posterior $rpgeo$ of \citet{bai21}. As most of the optically brightest stars did not have a {\it GAIA} parallax, their distance was taken from parallaxes listed in the new reduction of {\it Hipparcos} data \citep{van07} if $R_{plx}>5$. Considering only the stars in the western Galactic hemisphere (eROSITA\_DE half-sky) led to a final list of 832 stars, of which 170 were detected by eROSITA (see next section for details). It may be noted that some known \gc\ stars are found amongst the 170 detections. In fact, there are 10 known \gc\ and 2 known \gc\ candidates in the studied half-sky, but one object of each category was not considered in our Be star catalog (V767\,Cen was excluded because of its large $RUWE$, and HD\,120678 is not listed on the BeSS website). This implies that all known and available \gc\ stars and candidate, 9+1 in total, were in fact detected in the survey.

Table \ref{Targets} lists the detected objects with their properties. The spectral types were taken from Simbad or BeSS (if different, the most recent determination is used). "Early" and "late" classification in this paper corresponds to stars earlier than or equal to B3, and later than B3, respectively. The reddenings were calculated using {\sc stilism}\footnote{https://stilism.obspm.fr/} \citep{lal14} with the Simbad galactic coordinates and the known distances. In case the distances were larger than the maximum distance allowed by {\sc stilism}, then the reddenings were derived from the observed $B-V$(Simbad) and the intrinsic colors of the updated table of  \citet{mam13}\footnote{see {\scriptsize http://www.pas.rochester.edu/$\sim$emamajek/EEM\_dwarf\_UBVIJHK\_colors\_Teff.txt}} or of the table of \citet{mar06}. The reddenings $E(B-V)$ were then converted to interstellar absorbing columns using $N^{\rm ISM}_{\rm H}=6.12\times10^{21}\times E(B-V)$ \citep{gud12}. 

Bolometric luminosities were calculated in the usual way using the observed $V$ magnitudes (from Simbad\footnote{except for HD\,85860 for which there is no $V$  in Simbad, but one exists in BeSS.}), the derived reddenings (assuming $R_V=3.1$), the known distances, and bolometric corrections valid for the spectral types of the targets. The latter values were taken from the updated version of \citet{mam13} for dwarf stars or \citet{mar06} for giant and supergiant O-type stars. For supergiant B-type stars, the effective temperature scale of \citet{mar08} was used with the formula of \citet{ped20} to derive the corrections. Average values between supergiant and main sequence objects were then used for the giant B-type stars. For those calculations, in case the spectral types were simply "Be" (i.e. they are unknown), the most common Be spectral type, B2V, was assumed for the luminosity calculation; in case the luminosity class was not known, main sequence was assumed. 

\section{eROSITA}

eROSITA \citep[extended ROentgen Survey with an Imaging Telescope Array,][]{pre21} is the soft X-ray instrument onboard the SRG spacecraft \citep{srg}.
Its main goal is to perform an all-sky survey in the 0.2\,--\,10.0~keV energy range, the eROSITA all-sky survey (eRASS). The here analysed data is provided by the first four all-sky surveys with a duration of about 0.5~yr each, that were taken between December 2019 and December 2021. 

In survey mode the spacecraft rotates continuously and thereby during the eRASS execution each sky-location is observed multiple times by smoothly progressing and overlapping stripes that are scanned by the instrument. The basic survey geometry follows ecliptic coordinates with the rotation axis pointing roughly in the direction of the Sun and the ecliptic poles as survey poles, thus temporal source coverage and average exposure increase from the ecliptic plane towards the poles. With a scan rotation period of 4~h, a field of view (FOV) of 1.03$^{\circ}$ diameter and an average survey rate of 1$^{\circ}$/day, each position in the ecliptic is scanned on average about six times and receives an exposure of roughly 200~s per all-sky survey. The many technical constraints and operational details imposed on the survey strategy result in time-dependent modifications of the survey rate and thus the survey depth differs by about $\pm 25\,\%$ along the ecliptic plane. However, due to the high survey efficiency, no data gaps, i.e. unobserved regions, exist in the individual all-sky surveys. eRASS products are available for each individual all-sky survey as well as for the combined survey data. The eRASS1 data and catalogues are described in \cite{erass1}.

\subsection{eRASS catalogs \& counterparts}

We use standard source catalogs from the individual all-sky surveys (eRASS1, 2, 3, or 4) as well as the combined all-sky survey (eRASS:4). These catalogs are produced by the eROSITA data analysis pipeline that is based on the science analysis software (eSASS, \citealt{esass}); here we use the processing version `020'. Two independent detection runs are performed per data set, a single band detection run (1B, 0.2\,--\,2.3 keV) that is complemented by aperture photometry in several bands (0.2\,--\,0.5, 0.5\,--\,1.0, 1.0\,--\,2.0, 2.0\,--\,5.0 keV) and a three band detection (3B, 0.2\,--\,0.6, 0.6\,--\,2.3, 2.3\,--\,5.0 keV).

\begin{figure}
\includegraphics[width=\columnwidth]{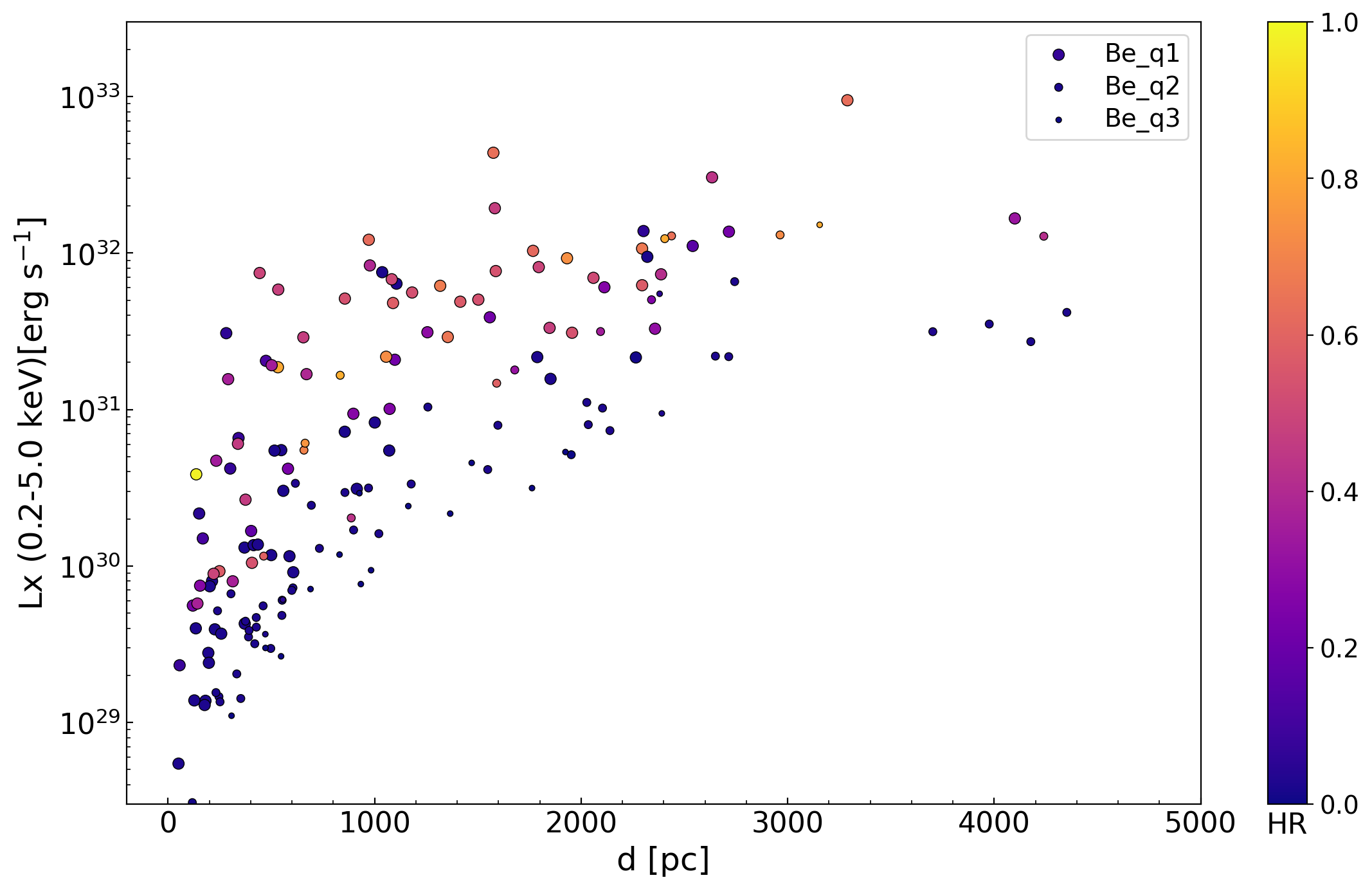}
\caption{eRASS detected Be stars (Lx vs. distance vs. spectral hardness) with q1, q2, q3 denoting detection quality; see text for details.}
\label{erass_be}
\end{figure}

The Be stars catalog is matched against the available eROSITA catalogs. We extracted matches if they are within 10\arcsec of the stellar position or if they are within two sigma positional error and within 20\arcsec radius, thereby accounting for systematic errors and uncertainties in the derived X-ray position. Our main identifications are based on the single band catalog from the combined eRASS:4 data, that is the most sensitive one. Other and individual eRASS catalogs are used as supplement and to investigate potential variability of the targets. Subsequent processing is based on X-ray data quality and detection likelihood as well as stellar properties like optical magnitudes. The X-ray luminosity vs. distance distribution of the Be stars selected for analysis is shown in Fig.\,\ref{erass_be}, color coding depicting spectral hardness ($ HR = 1 - F^{ISM\,cor}_{\rm X}(0.5-2.)/F^{ISM\,cor}_{\rm X}(0.5-5.)$, see more discussion below).

We define three grades of detection quality. A detection is denoted as `very good' (q1 in Fig. \ref{erass_be}), if a source has DET\_LIKE $\ge 10$ in the eRASS:4 catalog and is present in at least two eRASSn catalogs with an average DET\_LIKE $\ge 10$ in both detection runs, as `good' (q2), if a source has DET\_LIKE $\ge 8$ in an eRASS:4 catalog or if it is present in an eRASS:4 catalog and has an average DET\_LIKE $\ge 8$ from at least two eRASS catalogs and as `fair' (q3) if a source has DET\_LIKE $\ge 6$ in an eRASS:4 catalog or as average in at least two eRASSn catalogs. Sources detected below this threshold are not considered in the analysis.
This filter process ensures an overall high data quality and efficiently removes associations with spurious sources. To estimate the number of spurious identifications, we randomly shifted each catalogued Be star by a value uniformly distributed between 0.05 and 0.75 degree and the resulting sample was again cross-matched in the same way as before. This yields a remaining contamination of about $2 \pm 1$ detections in the `very good+good' combined as well as in the `fair' sample. In addition, we marked targets as potentially blended or confused if the X-ray source is detected as mildly extended or associated to more than one Be star counterpart.

eROSITA is prone to optical loading, i.e. bright sources produce fake X-rays in the detectors, predominantly at softer X-ray energies, which distorts the true source signal. As its strength mainly depends on magnitude, we removed very bright stars (V $< 2$~mag) from the sample and applied corrections for stars with V~$\le 5$~mag. For stars brighter than 4~mag, both the 0.2\,--\,0.6 keV and the 0.6\,--\,2.3 keV bands (or 0.2\,--\,0.5 keV and 0.5\,--\,1.0 keV in aperture photometry) are corrected, at 4\,--\,5 mag only the softest bands are corrected. The applied correction is primarily based on a study of early A~stars, that are intrinsically X-ray dark and optically bright. 
For our sample stars, we thus estimate and subtract an applicable offset from the measured rate of each energy band; if the detection remains significant, the star is kept with a corrected rate. This cleaning procedure eliminates optical contamination of sources quite efficiently, but reduces the sensitivity for optically bright sources with soft spectra.
We cross-checked our results with those from our spectral analysis and cross-matching of sources to catalog data like the RASS \citep[ROSAT all-sky survey,][]{vog99}. Detection quality, potential blends and optical contamination are tracked by a combined source quality flag in Table~\ref{Targets}.

In total, our initial matching procedure results in 195 detections. Filtering for optical contamination as well as likely erroneous or severely confused sources, 170 detections remain that are summarised in Table~\ref{Targets}. Note that the given count rates are corrected for vignetting and instrument performance and thus correspond to the full instrument on-axis equivalent values. Count rates were then transferred into fluxes via appropriate energy conversion factors (ECFs) derived from thermal plasma models \citep[APEC,][]{smi01} and eROSITA response matrices (ancillary response file, ARF, and redistribution matrix file, RMF). We applied a $kT =1$~keV plasma model and several two component models with temperatures in the range of 0.2 to 5.0~keV and find a good agreement within about 20\%. As the EFCs are applied band-wise, derived results are robust and overall independent of the very details of the true spectrum. 
Table \ref{stats} provides some statistics on detections: as could be expected, early-type and/or close Be stars have the largest detection rate.

\setcounter{table}{1}
\begin{table}
\scriptsize
\caption{Statistics on the sample ("u" is for unknown). For homogeneity, Simbad spectral types were used here for all stars. \label{stats}}
\begin{center}
  \begin{tabular}{lcccccccc}
    \hline
    Distance & \multicolumn{4}{c}{Full sample} & \multicolumn{4}{c}{Detections} \\
    (pc) & all & early & late & u & all & early & late & u\\
    \hline
all        & 832 & 382 & 365 & 85 & 170 (20\%) & 104 (27\%) & 50 (14\%) & 16 (19\%) \\
0--100     & 11 &  2 &  9 & 0 & 2  (18\%) &  0 (0\%)  &  2 (22\%) & 0        \\
100--250   & 50 & 17 & 33 & 0 & 22 (44\%) & 10 (59\%) & 12 (36\%) & 0        \\
250--500   & 95 & 47 & 46 & 2 & 33 (35\%) & 18 (38\%) & 15 (33\%) & 0        \\
500--1000  &226 & 84 &138 & 4 & 38 (17\%) & 23 (27\%) & 14 (10\%) & 1 (25\%) \\
1000--2000 &268 &132 &111 &25 & 39 (15\%) & 29 (22\%) &  5 (5\%)  & 5 (20\%) \\
2000--4000 &163 & 91 & 26 &46 & 30 (18\%) & 21 (23\%) &  2 (8\%)  & 7 (15\%) \\
4000--8000 & 19 &  9 &  2 & 8 &  6 (32\%) &  3 (33\%) &  0  (0\%) & 3 (38\%) \\
    \hline
  \end{tabular}
  \end{center}
\end{table}

\subsection{X-ray data analysis}
In addition to count rates, light curves and spectra generated with eSASS \texttt{srctool} from eROSITA event files, were also examined.

\subsubsection{Light curves and variability}

For the creation of X-ray light curves we use the 0.2\,--\,5.0\,keV energy band, where the survey is most sensitive for our targets. To study variability on longer timescales we bin them to the four individual eRASS all-sky surveys, i.e. data points are separated by about 0.5~yr each. In addition, for X-ray bright sources with count rates above 1~cts\,s$^{-1}$ light curves binned to the individual scans within each eRASS are created, thereby allowing to also sample timescales of hours up to days. To this aim, the source photons are extracted from circular regions around the respective X-ray positions; the background was taken from larger annuli. The standard parameter are $r_{\rm src}=0.01^{\circ}$ and $r_{\rm bg}=0.02^{\circ}-0.04^{\circ}$, but regions were adapted if required. As some blends from close-by stars are not ruled out, we inspected the data for potential contamination. Overall, we find that the main target is the dominant X-ray source, but in dense environments like Orion some residual contribution from neighboring sources might be present. In one case (15\,Mon) we identified bright, flaring emission from a nearby star that appeared as a transient second source in several scans in eRASS1, here time filtering was applied to clean the data.

\begin{figure}
\includegraphics[width=\columnwidth]{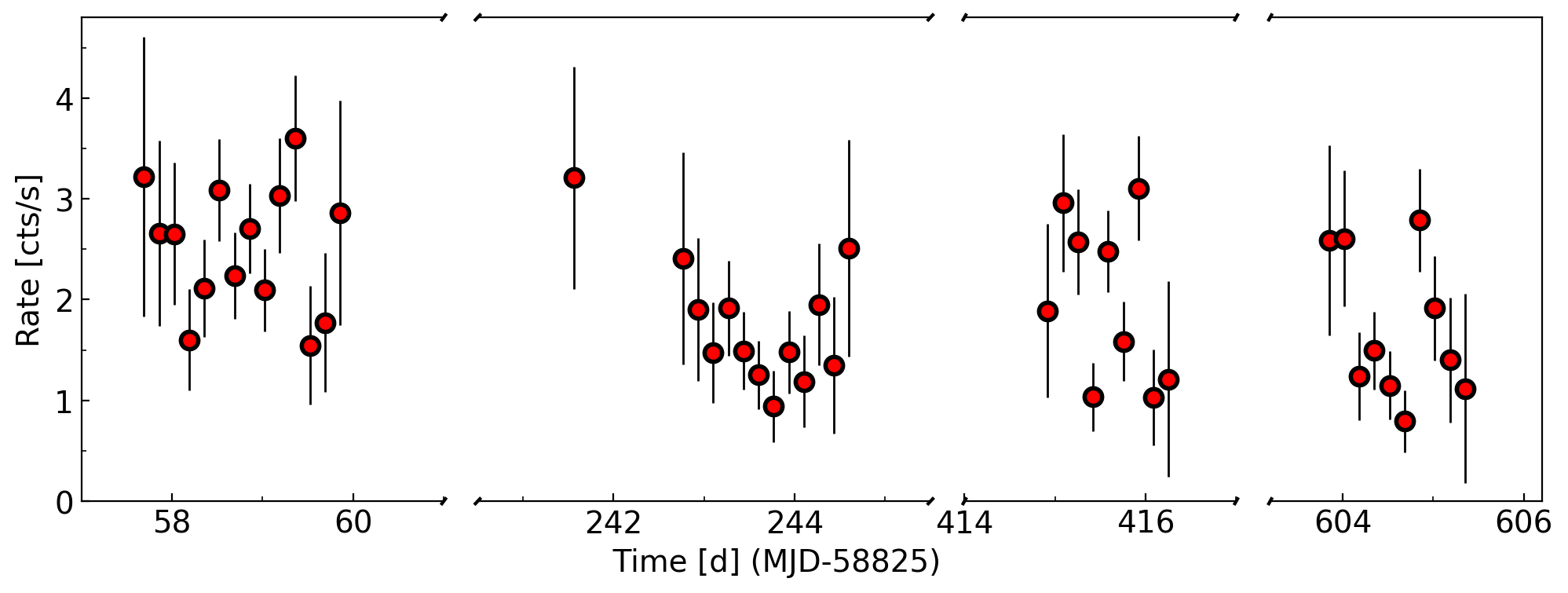}
\includegraphics[width=\columnwidth]{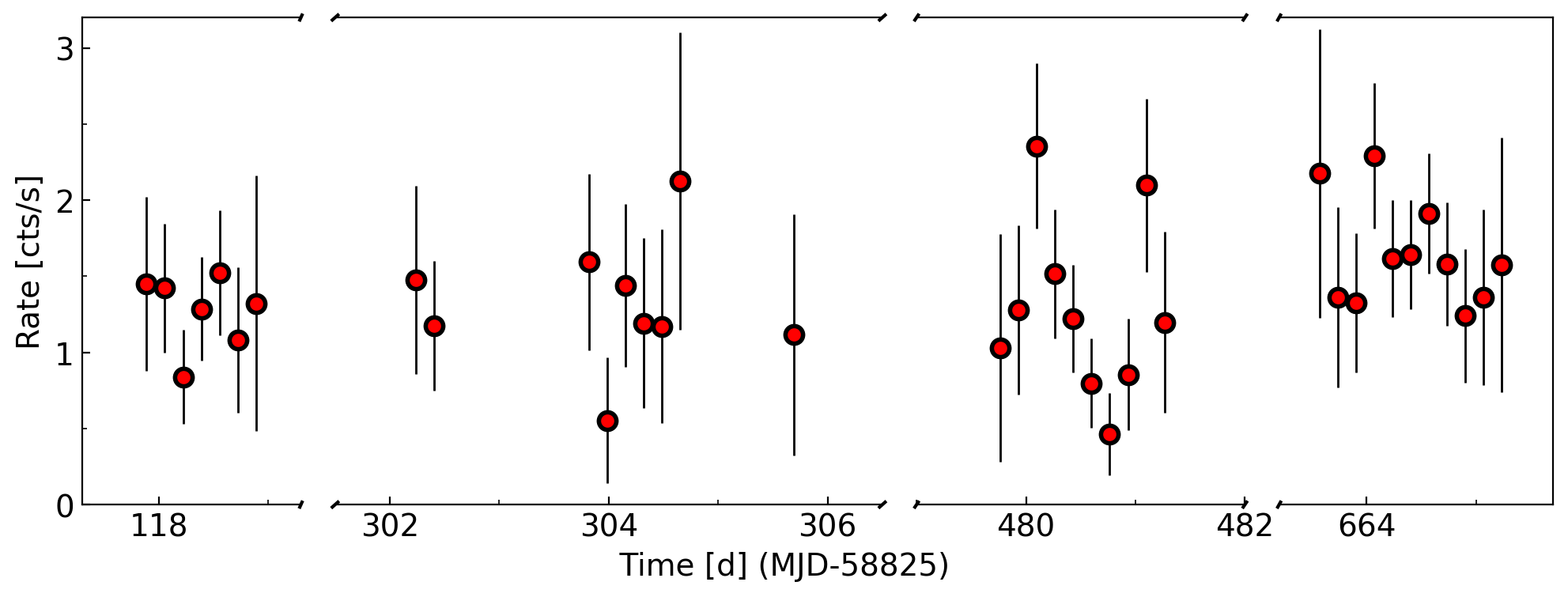}
\caption{eRASS1-4 scan resolved light curves of BZ~Cru (top) and HD~42054 (bottom), a $\gamma$\,Cas analog and a $\gamma$\,Cas candidate source.}
\label{erasslc0}
\end{figure}

All light curves are corrected for vignetting and instrument performance. If present, optical loading is taken into account for subsequent analysis steps. For the studied stars these light curves often appear quite erratic, however coherent variability on timescales of hours up to a day is also seen at other times e.g. for several $\gamma$\,Cas like objects as shown in Fig. \ref{erasslc0}. Comparable short term variability is likely also present in several other sources, but could not be studied in detail for a large sample due to the low number of counts obtained during the short duration of the survey-scans.

\begin{figure}
\includegraphics[width=\columnwidth]{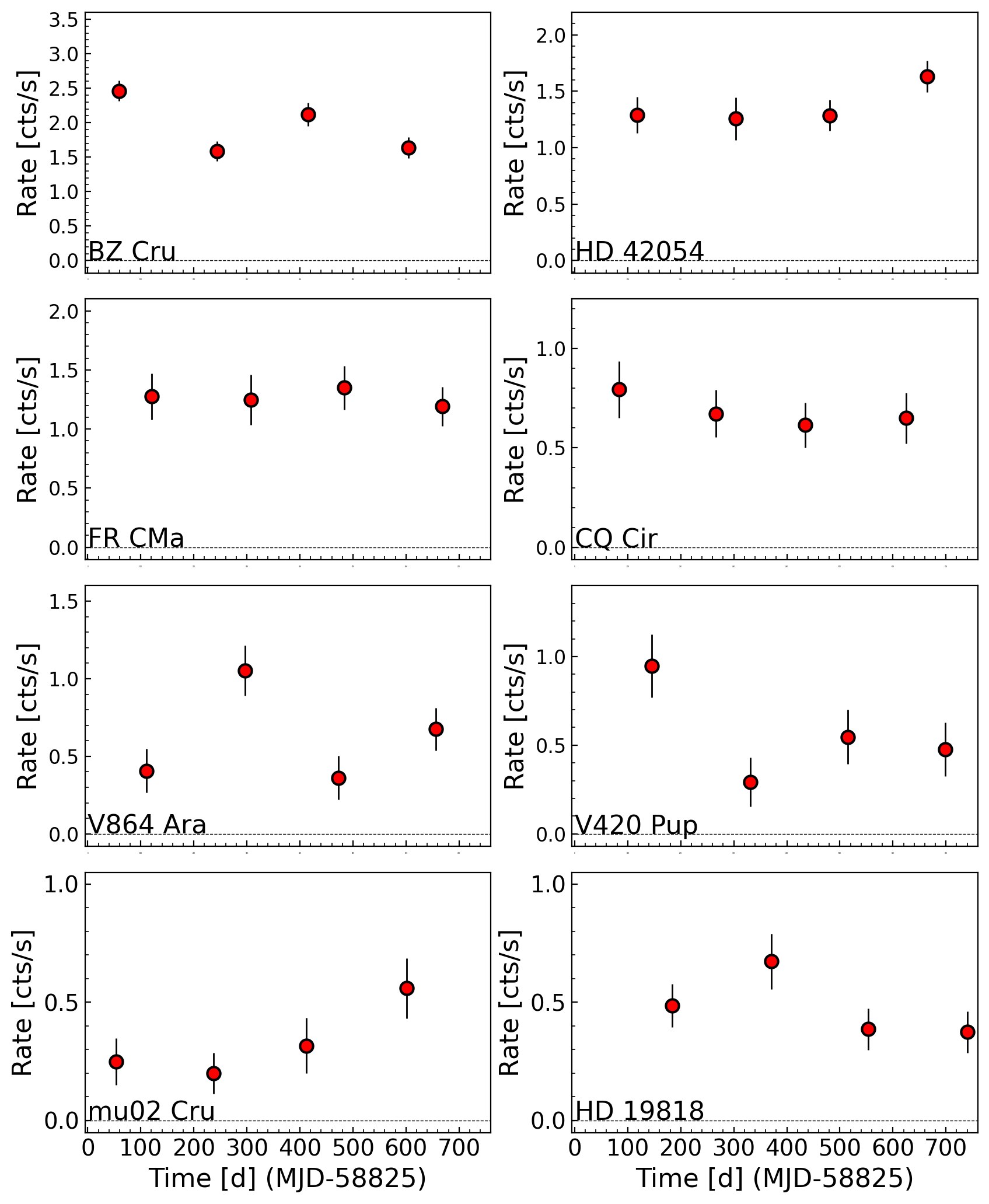}
\caption{eRASS1-4 survey binned light curves of Be stars in the 0.2\,--\,5.0~keV energy band. The upper four are $\gamma$\,Cas like sources or candidates and mostly of earlier type (B0.5-B5), the lower four ones are other Be stars and mostly of later types (B3-B9.5).}
\label{erasslc1}
\end{figure}

When inspecting the generated light curves, some variability is also found on longer timescales, i.e. between the four individual eRASS all-sky scans. Nevertheless, overall the investigated Be stars are found to be quite stable X-ray emitters over the two years timescale and their light curves are mostly smooth or show mild variability. The X-ray flux changes are within a factor of two to three, typically lower. Large outbursts or strong flares were not observed in the eRASS data of the studied stars. Example eRASS light curves of some X-ray brighter sources are shown in Fig.\,\ref{erasslc1}.

To quantify variability in the Be star sample, we also use the count rates from eRASS1-4 as given in the source catalogs from the 0.2\,--\,2.3~keV detection run. We study sources that are detected in more than half of the all-sky scans and calculate their maximum to minimum eRASSn rate ratio.
Variability at a minimum significance of $3 \sigma$ is detected in V420\,Pup (ratio of $3.41 \pm 0.80$), V864\,Ara ($2.78 \pm 0.54$), $\mu^{02}$\,Cru ($2.47 \pm 0.46$), HD\,19818 ($1.73 \pm  0.24$), BZ\,Cru ($1.52 \pm 0.11$). In addition, variability above $2 \sigma$ is found in CD--29\,5159 ($3.93 \pm 1.42$), MQ\,TrA ($ 2.74 \pm 0.65 $), V863\,Cen ($2.55 \pm 0.72$), HD\,79778 ($2.28 \pm 0.67 $), CK\,Cir ($2.11 \pm 0.48 $). 
There are also a few sources with an average eRASS:4 count rate above 0.1 cts\,s$^{-1}$, that remained undetected in at least one all-sky survey. These are HD\,308829 (count rate of 0.15 cts\,s$^{-1}$), V1083\,Sco (0.12 cts\,s$^{-1}$), HD\,161774 (0.10 cts\,s$^{-1}$) and HD\,141926 (0.10 cts\,s$^{-1}$), indicating variability of at least a factor of a few given average survey sensitivity.

\subsubsection{Spectral analysis}
\begin{figure}
\includegraphics[width=\columnwidth]{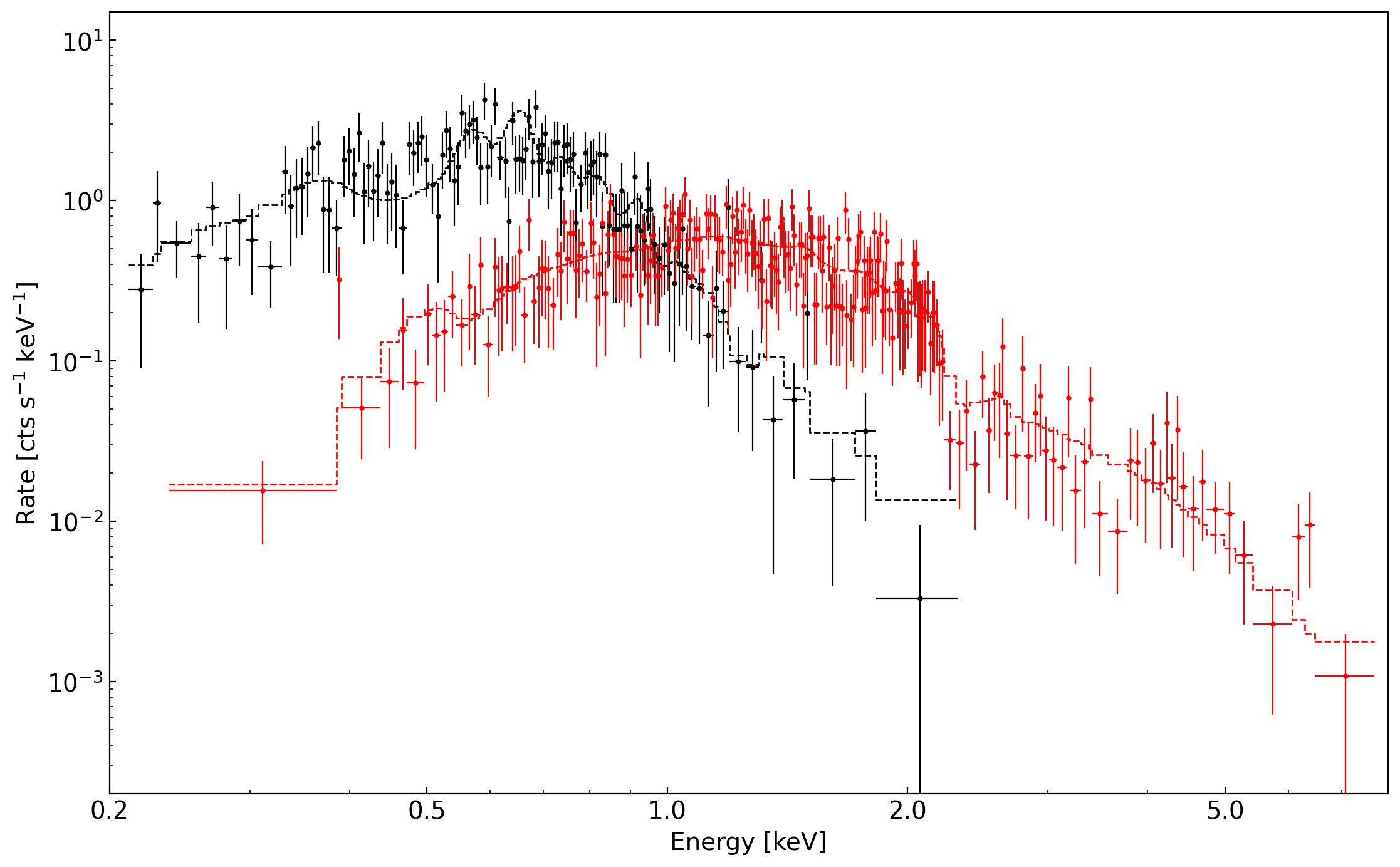}
\caption{Example spectra obtained from eRASS:4 data, 15 Mon (black) and BZ Cru (red) with best-fit thermal plasma model.}
\label{erasssp1}
\end{figure}

Stellar spectra were obtained from the combined eRASS data in the same regions as the light curves and if at least 200 counts were recorded overall, which is the case of about twenty sources. Such spectra then describe average stellar properties in the 0.2\,--\,10.0~keV energy range during the survey (for their variability, see previous paragraphs). Source specific ancillary response files (ARFs) were calculated to allow the spectral analysis. The eRASS:4 spectra from two X-ray brighter Be stars are shown in Fig.\,\ref{erasssp1}.

We have modelled the spectra as in \citet{naz18}, i.e. using an optically thin thermal $apec$ model with both interstellar and circumstellar absorption. Several $apec$ were only used if a single component was not sufficient to achieve a good fitting. The interstellar absorbing column was fixed to that determined from the reddening (see Section 2) and spectra were binned to reach a minimum of 10 counts per bin. In one case, 43\,Ori, the large uncertainties prohibited any meaningful fitting. In five other cases (the optically bright stars $\zeta$\,Tau, $\delta$\,Cen, $\mu$\,Cen, $\delta$\,Sco, and $\gamma$\,Ara), an increase of counts towards low energies is detected. This is due to optical loading. To model it, a power law of initial index 6.3 as derived from the above mentioned study of optically bright A-type stars was added to the thermal model. It usually provides good results, except for $\delta$\,Sco, for which the intrinsic emission is so soft that it cannot be clearly disentangled from the optical loading. The results from the spectral fittings are summarized in Table \ref{specfit}. They usually are coherent with those reported in \citet{naz18}, although the errors are quite large in our data and some (limited) brightness variability is detected in some cases. 

\begin{table*}
  \scriptsize
  \caption{Results of the spectral fittings.  \label{specfit}}
  \begin{tabular}{lcccccccccc}
    \hline
Name  & $N^{\rm ISM}_{\rm H}$ & $N_{\rm H}$ & $kT_1$ & $norm_1$ & $kT_2$ & $norm_2$ & $\chi^2$/dof & \multicolumn{3}{c}{$F_{\rm X}^{obs}$(erg\,cm$^{-2}$\,s$^{-1}$)}  \\
     &($10^{22}$\,cm$^{-2}$) &($10^{22}$\,cm$^{-2}$) & (keV) & (cm$^{-5}$) &(keV) & ($10^{-3}$\,cm$^{-5}$) & & 0.5-5.\,keV & 0.5-2.\,keV & 2.-10.\,keV \\
\hline
HD\,19818	&0.010 &0.014$\pm$0.016& 0.30$\pm$0.03 & (9.65$\pm$1.90)e-5 & 1.51$\pm$0.25 & (1.81$\pm$0.39)e-4 & 36.93/32	& (3.25$\pm$0.26)e-13 & (2.73$\pm$0.17)e-13 & (5.96$\pm$1.34)e-14 \\
$\zeta$\,Tau	&0.019 & 11.8$\pm$4.2  & 64.0$\pm$59.3 & (1.64$\pm$0.38)e-2 &               &                    & 22.28/26	& (3.34$\pm$5.35)e-12 & (0.14$\pm$4.42)e-13 & (1.29$\pm$1.30)e-11 \\
HD\,42054	&0.0055&0.014$\pm$0.009& 0.73$\pm$0.09 & (7.60$\pm$1.76)e-5 & 3.66$\pm$0.89 & (1.15$\pm$0.09)e-3 & 50.34/60	& (1.46$\pm$0.09)e-12 & (8.72$\pm$0.36)e-13 & (9.32$\pm$2.17)e-13 \\
FR\,CMa	        &0.054 & 0.12$\pm$0.04 & 0.65$\pm$0.27 & (3.87$\pm$3.12)e-5 & 5.27$\pm$2.13 & (1.80$\pm$0.16)e-3 & 39.42/37	& (1.89$\pm$0.19)e-12 & (8.67$\pm$0.54)e-13 & (1.85$\pm$0.43)e-12 \\
15\,Mon	        &0.0067&0.021$\pm$0.007& 0.24$\pm$0.01 & (1.61$\pm$0.13)e-3 & 1.97$\pm$0.67 & (3.27$\pm$2.16)e-4 & 64.26/53	& (1.76$\pm$0.16)e-12 & (1.64$\pm$0.11)e-12 & (1.47$\pm$1.37)e-13 \\
V420\,Pup	&0.036 &0.000$\pm$0.021& 1.02$\pm$0.17 & (6.30$\pm$4.25)e-5 & 4.45$\pm$8.30 & (4.44$\pm$0.78)e-4 & 12.7/13	& (6.14$\pm$1.00)e-13 & (3.60$\pm$0.45)e-13 & (4.25$\pm$2.80)e-13 \\
HD\,97253	&0.27  & 0.00$\pm$0.07 & 0.74$\pm$0.06 & (1.34$\pm$0.12)e-4 &               &                    & 28.91/18	& (1.68$\pm$0.23)e-13 & (1.57$\pm$0.18)e-13 & (1.10$\pm$0.29)e-14 \\
$\delta$\,Cen	&0.0067& 0.36$\pm$0.26 & 0.26$\pm$0.07 & (2.56$\pm$6.51)e-4 &               &                    & 53.89/45	& (7.72$\pm$3.90)e-14 & (7.70$\pm$3.90)e-14 & (2.25$\pm$1.38)e-16 \\
BZ\,Cru	        &0.20  & 0 (fixed)     & 64.0$\pm$34.4 & (5.35$\pm$0.16)e-3 &               &                    & 106.78/105	& (4.13$\pm$1.12)e-12 & (1.34$\pm$0.40)e-12 & (6.46$\pm$1.45)e-12 \\
$\mu^{02}$\,Cru	&0.010 &0.000$\pm$0.014&0.23$\pm$0.03  & (5.22$\pm$1.48)e-5 & 1.07$\pm$0.09 & (8.20$\pm$1.30)e-5 & 14.8/13	& (1.93$\pm$0.22)e-13 & (1.78$\pm$0.20)e-13 & (1.61$\pm$0.37)e-14 \\
HD\,119682	&0.18  & 0 (fixed)     & 0.10$\pm$0.04 & (0.30$\pm$6.46)e-3 & 64.0$\pm$41.6 & (9.95$\pm$2.41)e-4 & 19.9/19	& (7.85$\pm$8.65)e-13 & (2.67$\pm$2.60)e-13 & (1.20$\pm$1.30)e-12 \\
$\mu$\,Cen	&0.008 & 1.01$\pm$0.14 & 0.19$\pm$0.05 & (1.00$\pm$1.42)e-2 &               &                    & 22.94/18	& (1.81$\pm$1.02)e-13 & (1.81$\pm$1.16)e-13 & (5.59$\pm$3.57)e-16 \\
CQ\,Cir	        &0.25  & 0.54$\pm$0.10 &64.0$\pm$44.7 &(3.31$\pm$0.72)e-3 &  &  & 33.25/32	& (2.07$\pm$1.70)e-12 & (4.53$\pm$4.00)e-13 & (3.87$\pm$4.00)e-12 \\
V1075\,Sco	&0.17  & 0.00$\pm$0.02 & 0.26$\pm$0.01 & (8.02$\pm$1.19)e-4 &               &                    & 26.89/22	& (4.51$\pm$0.70)e-13 & (4.50$\pm$0.85)e-13 & (8.14$\pm$2.24)e-16 \\
$\gamma$\,Ara	&0.061 & 0.29$\pm$0.14 & 0.62$\pm$0.07 & (3.76$\pm$1.65)e-4 &               &                    & 41.93/33	& (3.72$\pm$0.46)e-13 & (3.55$\pm$0.46)e-13 & (1.80$\pm$0.43)e-14 \\
V750\,Ara	&0.081 & 0.72$\pm$0.48 & 0.55$\pm$0.31 & (1.87$\pm$4.93)e-4 & 6.50$\pm$4.48 & (1.44$\pm$0.24)e-3 & 19.9/16	& (1.19$\pm$0.30)e-12 & (3.84$\pm$1.06)e-13 & (1.58$\pm$1.00)e-12 \\
V864\,Ara	&0.035 & 0 (fixed)     & 0.72$\pm$0.11 & (7.64$\pm$1.67)e-5 & 3.86$\pm$2.03 & (4.11$\pm$0.60)e-4 & 22.98/22	& (6.18$\pm$0.71)e-13 & (3.99$\pm$0.50)e-13 & (3.52$\pm$1.60)e-13 \\
$\beta$\,Scl	&0.0006&0.068$\pm$0.036&0.084$\pm$0.025& (1.49$\pm$2.44)e-3 & 0.88$\pm$0.05 & (1.97$\pm$0.24)e-4 & 34.59/15	& (4.05$\pm$0.66)e-13 & (3.80$\pm$0.60)e-13 & (2.59$\pm$0.50)e-14 \\
\hline
CD--29\,5159	&0.48 & 0.51$\pm$0.08 & 64.0$\pm$19.9 & (1.50$\pm$0.11)e-3 &	            &                    & 215.47/189 & (8.93$\pm$1.43)e-13 & (1.76$\pm$0.44)e-13 & (1.74$\pm$0.23)e-12 \\
HD\,302798	&0.32 & 0.14$\pm$0.11 & 4.84$\pm$2.00 & (1.86$\pm$0.19)e-4 &	            &                    & 51.2/43    & (1.58$\pm$0.13)e-13 & (5.83$\pm$0.43)e-14 & (1.77$\pm$0.31)e-13 \\
HD\,305560  &0.36 & 3.95$\pm$0.26 & 64.0$\pm$4.58 & (2.69$\pm$0.10)e-1 &                &                    & 190.85/155 & (9.98$\pm$2.00)e-11 & (5.15$\pm$1.80)e-12 & (2.69$\pm$0.48)e-11 \\
$\mu^{02}$\,Cru	&0.010 & 0.10$\pm$0.06 & 0.24$\pm$0.01 & (1.32$\pm$0.55)e-4 & 1.12$\pm$0.05 & (6.16$\pm$0.50)e-5 & 262.29/214 & (1.70$\pm$0.07)e-13 & (1.58$\pm$0.07)e-13 & (1.28$\pm$0.13)e-14 \\
V955\,Cen	&0.23 & 0.00$\pm$0.03 & 0.36$\pm$0.08 & (2.13$\pm$1.17)e-5 & 4.70$\pm$0.34 & (3.15$\pm$0.09)e-4 & 576.09/529 & (3.20$\pm$0.10)e-13 & (1.48$\pm$0.04)e-13 & (3.00$\pm$0.16)e-13 \\
\hline
  \end{tabular}
  
{\scriptsize Models are of the form $tbabs\times phabs \times \sum apec$ as in \citet{naz18}. The errors correspond to 1$\sigma$ errors and, if asymmetric, the largest value is provided. Top part yields the results for eROSITA spectra while the bottom part corresponds to comparison XMM (ObsID=0860650501, 0861590101, 0884550101), Chandra (ObsID=22454), and Swift data (ObsID=00614193) - see text for details. }
\end{table*}

\section{Results}
Massive stars with spectral type O and early B possess strong stellar winds which are radiatively driven. As this line-driving process is unstable, shocks arise in the winds, giving rise to hot plasma \citep{fel97}. There is an interplay between emission and absorption in such winds which gives rise to the well-known relationship $L_{\rm X}/L_{\rm BOL}\sim 10^{-7}$ \citep{owo99,owo13}. Therefore, the X-ray luminosity quoted here is the luminosity calculated by correcting the observed fluxes for interstellar absorption only (i.e. not for circumstellar absorption), and it is the one listed in Table \ref{Targets}. In addition, this X-ray emission is quite soft ($kT=0.2-0.6$\,keV, see e.g. \citealt{naz09}). In this framework, late-type B-stars are not expected to be X-ray bright as their winds are too tenuous. Deviations from this "canonical" relationship are however detected in case of strong magnetism \citep{naz14} or of colliding winds (see e.g. the review in \citealt{rau16}). In both cases, the X-ray luminosity increases, and often the hardness too, although rarely beyond $\log(L_{\rm X}/L_{\rm BOL})=-6.2$ and $kT$=2\,keV. Very hard and bright emissions are however spotted for \gc\ stars and, even more extreme, for HMXBs (although it must be remembered that known X-ray binaries were discarded from our stellar catalog so should not appear amongst detections). Finally, if a massive star is paired with a pre-main sequence star (PMS), this PMS can also be a source of X-rays, with luminosities up to $10^{31}$\,erg\,s$^{-1}$ and $kT\sim 2$\,keV during flaring.  

Various X-ray properties can thus be expected, depending on spectral type, multiplicity, magnetism,... In this study, we rely on the usual three main diagnostics: The X-ray luminosity, the X-ray to bolometric luminosity ratio, and the hardness of the X-ray emission, as they have proven to be quite sensitive markers. In this paper as usually in massive star papers, we use fluxes corrected only for interstellar absorption to derive these diagnostics (see above). We also use here a high-energy limit of 5\,keV, rather than the usual 10\,keV of XMM or Chandra papers, because of the low sensitivity of the eROSITA survey beyond that energy which, coupled to the generally low number of photons emitted by massive stars beyond that energy, makes the count rate of the survey rather low, hence unreliable, in the 5.\,--\,10.\,keV range. Finally, we define the hardness ratio $HR$ as the fraction of the emission made in the 2.\,--\,5.\,keV range. It is calculated as $1-F^{ISM\,cor}_{\rm X}(0.5-2.)/F^{ISM\,cor}_{\rm X}(0.5-5.)$, which is zero for a very soft source and one for a very hard one. It relies on the soft (0.5\,--\,2.\,keV) flux rather than the hard (2.\,--\,5.\,keV) one as the soft flux generally is much better defined thanks to a larger number of counts in that range.

\begin{figure}
\includegraphics[width=\columnwidth]{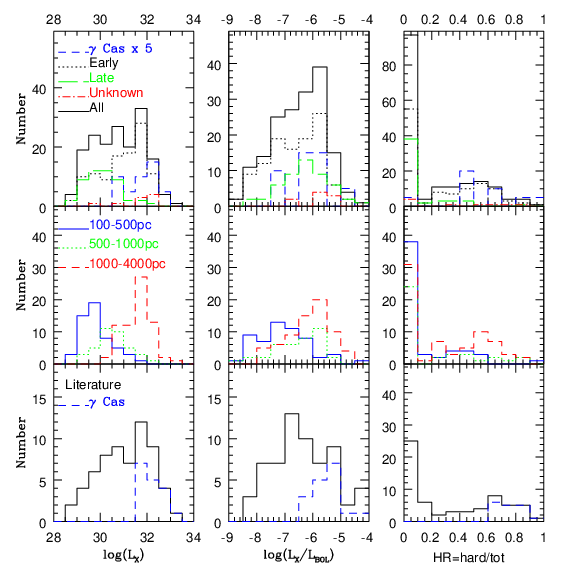}
\caption{Histograms of X-ray luminosities (in 0.5\,--\,5.\,keV, after correction for interstellar absorption), the X-ray to bolometric luminosity ratios, and hardnesses of the X-ray emissions of Be stars (defined as ratios of the hard, 2.\,--\,5.\,keV and total, 0.5\,--\,5.\,keV fluxes). Top and middle panels provide histograms for the targets studied by eROSITA, split by spectral type and distance, respectively. To ease comparison, the numbers associated to the known \gc\ stars were multiplied by a factor of five. The bottom panels provide histograms for Be samples previously published in literature (see text for details). Note that for the latter case, the upper limit for the high-energy range is 10\,keV rather than the 5\,keV adopted for eROSITA.}
\label{histo}
\end{figure}

Figure \ref{histo} provides the histograms of the three main diagnostics for our sample. It shows how they change depending on distance or spectral type of the targets. As could be expected, only the bright (and hard) X-ray sources are detected from afar. In addition, with 90\% of the X-ray luminosities below 10$^{31}$\,erg\,s$^{-1}$, the late-type stars appear less luminous than the early-type objects. Finally, it must be noted that most detections correspond to soft sources as they possess very low hardness ratios: values below 0.1 are found for 57\% of the whole sample (97/170), 50\% of the early-type stars (55/111), 76\% of the late-type stars (38/50), 44\% of the stars with unknown type (4/9), 69\% of stars within 100\,--\,500\,pc (38/55), 63\% of stars within 500\,--\,1000\,pc (24/38) and 45\% of stars within 1000\,--\,4000\,pc (31/69). Stars below 100\,pc or more distant than 4000\,pc correspond to only 5\% of detected objects and only 4\% amongst the 832 known Be stars of the eROSITA\_DE half-sky. They suffer from small numbers statistics and high incompleteness, so they are not considered in these distance statistics. Note that there is one known \gc\ star displaying such a low value of $HR$: PZ\,Gem. This is not surprising as this star has been reported to show a reduced hard X-ray emission as the optical emission from its disk decreased \citep{rau18}: the eROSITA data simply confirm the continuation of that low-emission phase. 

Figure \ref{histo} also compares the results for the eROSITA sample to 61 Be stars whose spectral properties have been recently published in literature: 47 stars from \citet{naz18}, 13 stars from \citet{naz20} and one star (HD\,194335) from \citet{naz22be} - the other ones in the latter paper were already published in the previous articles (59\,Cyg, V750\,Ara) or are somewhat uncertain ($\zeta$\,Tau). While the covered ranges are similar, a larger fraction of hard sources (only 40\% of sources have $HR<0.1$ in that sample) as well as of sources with high $\log(L_{\rm X}/L_{\rm BOL})$ ratios ($>5.5$) may be noted. This is not surprising since these non-random samples often corresponded to specific observing requests, not to a general survey as is the case of our eROSITA sample: biases were thus expected.

\begin{figure}
\includegraphics[width=\columnwidth]{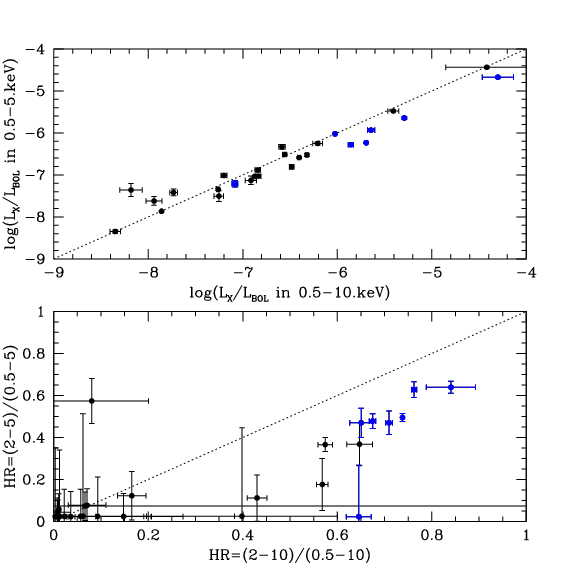}
\caption{Comparison of the X-ray luminosities and hardness ratios derived from eROSITA data in the 0.5--5.\,keV range with those previously published for the same targets in the 0.5--10.\,keV range \citep{naz18,naz20}. Symbols in blue are used for the known \gc\ stars.}
\label{check}
\end{figure}

Several checks were performed to assess the validity of our results and to ascertain their interpretation. First, we compared the $L_{\rm X}/L_{\rm BOL}$ ratios and the hardness ratios derived from the spectral fits (Table \ref{specfit}) and from the counts (Table \ref{Targets}). A good agreement is found, although the values are more uncertain in the former case. Second, we compared the same parameters, all derived from spectral fits but with an upper limit of 5\,keV or of 10\,keV. Again, a good agreement is found: the values are slightly larger if the upper limit is larger, as expected, but the difference remains small. Finally, we compared values found in our eROSITA survey using counts with those published previously for the same stars using spectral fitting (23 cases from \citealt{naz18}, excluding faint detections without spectra and a dubious identification, and 4 from \citealt{naz20} - see Fig. \ref{check}). The difference in upper limit (5 vs 10\,keV) makes the literature values larger, but the values are clearly proportional, within errors, and the difference remains small (about 0.2\,dex in $\log(L_{\rm X}/L_{\rm BOL})$ and about 0.2 in $HR$). 

\begin{figure*}
\includegraphics[width=8cm]{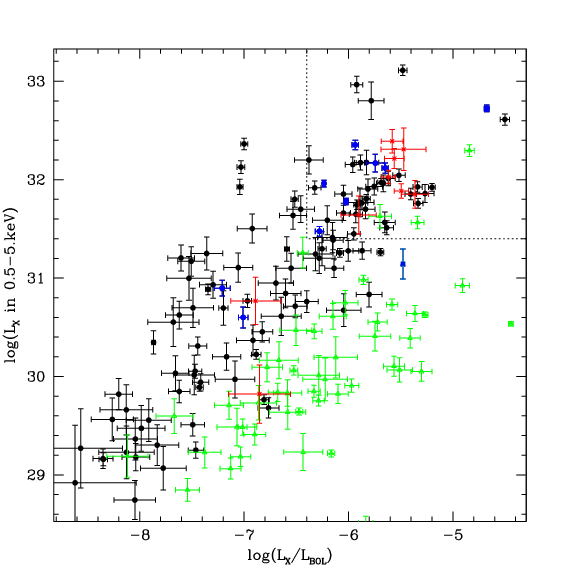}
\includegraphics[width=8cm]{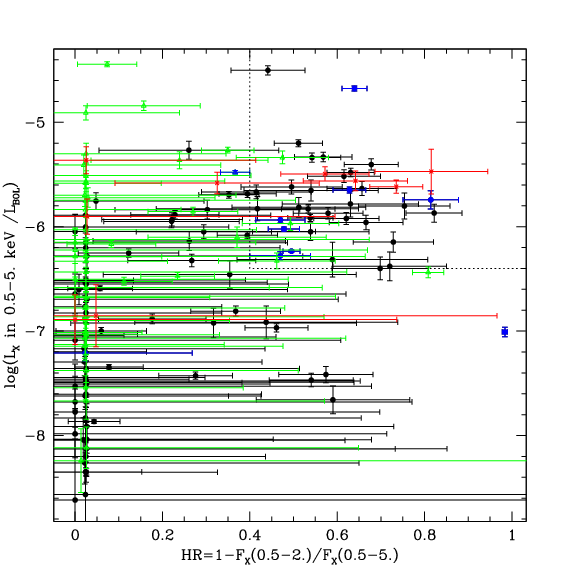}
\caption{Two by two comparison of X-ray luminosities (in 0.5--5.\,keV, after correction for interstellar absorption), the X-ray to bolometric luminosity ratios, and hardnesses of the X-ray emissions of Be stars (defined as ratios of the hard, 2.--5.\,keV and total, 0.5--5.\,keV fluxes). Early-type stars are shown with black dots, late-type stars with green triangles, stars of unknown types with red crosses, and known \gc\ stars and candidate as blue dot and blue triangle, respectively. 
}
\label{lxlb}
\end{figure*}

Figure \ref{lxlb} compares the X-ray luminosities to the X-ray to bolometric luminosity ratios of our targets and these ratios to their hardness ratios. For the same X-ray to bolometric luminosity ratio, the late-type stars appear at lower X-ray luminosities than the early-type objects. In addition, few late-type stars display high X-ray to bolometric luminosity ratios. Apart from that, no obvious clear-cut is apparent, e.g. stars at the highest luminosities and highest X-ray to bolometric luminosity ratios do not form a clearly separated group. Rather, a continuous trend towards higher values for both parameters is detected. 

The situation appears different for hardness ratios. The histograms revealed that a majority of sources have low hardness ratios ($HR<0.1$, Fig. \ref{histo}). Some sources also have very uncertain ratios, making them fully compatible with zero: they cannot be said to have a significant hardness. However, Figure \ref{lxlb} further shows that the sources with securely large ratios have, on average, larger X-ray to bolometric luminosity ratios. Indeed, sources with hardness ratios statistically larger than zero (i.e. $HR/\sigma_{HR}>1$) have average $L_{\rm X}(0.5-5.\,{\rm keV})= 1.1 \times 10^{32}$\,erg\,s$^{-1}$, average $\log(L_{\rm X}/L_{\rm BOL})=-5.9$, and average $HR=0.48$ while the other sources display averages of $0.13\times 10^{32}$\,erg\,s$^{-1}$, --6.9, and 0.03, respectively. Rather adopting a cut in $HR$ values, one gets: average $L_{\rm X}(0.5-5.\,{\rm keV})= 0.14$ or $1.1 \times 10^{32}$\,erg\,s$^{-1}$, average $\log(L_{\rm X}/L_{\rm BOL})=-6.9$ or $-6.0$, and average $HR=0.03$ or 0.50 for sources with $HR$ below or above 0.2, respectively. There thus seems to be a group of sources which are significantly hard and brighter at the same time.  Fixing a limit may seem quite difficult, in view of the dispersion and large error bars on $HR$, but the comparison done with literature values (see above) may help. The published criteria for identifying the bright and hard \gc\ objects amongst Be stars were: $\log(L_{\rm X}\,in\, 0.5-10.\,{\rm keV})$ in 31.6--33.2, $\log(L_{\rm X}\,in\, 0.5-10.\,{\rm keV}/L_{\rm BOL})$ between --6.2 and --4, $kT >5$\, keV, hardness defined as the ratio between hard (2.\,--\,10.\,keV) and soft (0.5\,--\,2.\,keV) fluxes larger than 1.6 (which corresponds to a fraction of hard-to-total flux, as used in this paper, $>0.6$), and $L_{\rm X}(2.-10.\,{\rm keV}) > 10^{31}$\,erg\,s \citep{naz18,naz20}. In this paper, the upper limit in energy is set at 5\,keV but the above comparison suggests that a change by 0.2 of the limits may be needed, i.e. $\log(L_{\rm X}\,in\, 0.5-5.\,{\rm keV})>31.4$, $\log(L_{\rm X}\,in\, 0.5-5.\,{\rm keV}/L_{\rm BOL})>-6.4$, $HR>0.4$. These limits are shown as dotted lines in Fig. \ref{lxlb}. 

\begin{figure}
\includegraphics[width=8cm]{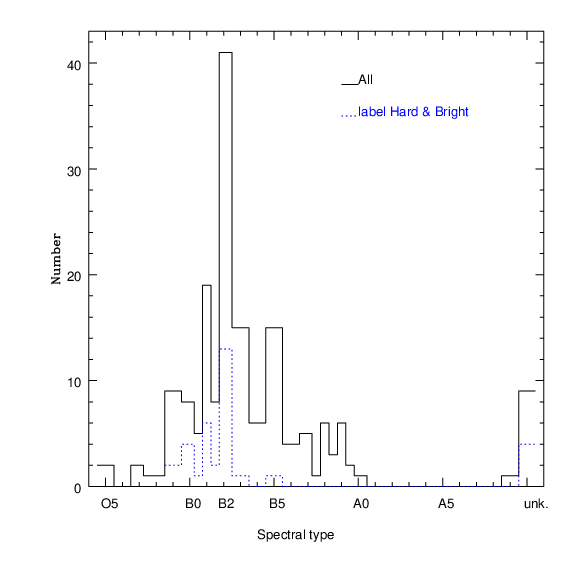}
\caption{Histogram of the spectral types amongst detected sources and amongst the bright and hard cases. }
\label{histo2}
\end{figure}

There are 34 sources with such properties amongst the whole sample (which corresponds to 20\% of the 170 detections). Splitting by type or distance, this leads to fractions of 26\% of the early-type stars (29/111), 2\% of the late-type stars (1/50), 44\% of the stars with unknown type (4/9), 2\% of stars within 100\,--,500\,pc (1/55), 13\% of stars within 500\,--\,1000\,pc (5/38) and 36\% of stars within 1000\,--\,4000\,pc (25/69). The increase of the incidence with distance is unsurprising as the harder sources are on average brighter and only bright sources will be detected from afar, skewing the distribution. In parallel, as already suggested before \citep{naz18}, the Be stars with bright and hard X-ray emission are preferentially found amongst early-type stars (see also Fig. \ref{histo2}). Indeed, only one late-type Be star is found to fulfill the criteria. To get an idea of the incidence of such a hard X-ray emission, we can thus restrict ourselves to early-type Be stars. However, one must also consider the distance effect. At large distances, bright X-ray sources such as those will be detected while the bulk of the sources, which are fainter, will not. Therefore, we restrict the sample to early-type stars in 100\,--\,1000\,pc, where the completeness is best and extremely small numbers (e.g. 1) are avoided: the derived fraction of bright and hard objects then is 12\% (6/51). Considering only Poisson errors, the 1$\sigma$ uncertainty on this fraction should be 5\%. This agrees well with some rough estimates of the \gc\ incidence rates made from the previous samples of Be stars (10--20\%, \citealt{naz22be}). It confirms that the emission of hard X-rays concerns a non-negligible fraction of Be stars. Finally, it may be noted that two of the known \gc\ objects do not fulfill the above criteria: PZ\,Gem which is in faint and soft state for the moment (see above) and $\zeta$\,Tau which, due to its extreme absorption, has a decreased X-ray luminosity although its spectral properties make it belonging to the \gc\ category without doubt (see Table \ref{specfit} and \citealt{naz22be}). 

The above criteria may be considered as quite strict. However, considering the current large values of the errors on the $HR$, it is difficult to lower the $HR$ limit further without adding potential interlopers. Nevertheless, the 18 objects not fulfilling all criteria but having at least $\log(L_{\rm X})>31$, $\log(L_{\rm X}/L_{\rm BOL})>-7$, and $HR>0.2$ are certainly interesting cases to further investigate. Indeed, if confirmed, they appear in-between the clearly soft objects and the clearly hard ones. In the context of the \gc\ phenomenon, they could either be objects in a short transition phase between those states or more constant objects but with a lower amplitude of the phenomenon causing the hard X-ray emission. In both cases, a better understanding of their properties might provide very important constraints, notably on the \gc\ phenomenon.

As a final check on the validity of the chosen criteria, a search for XMM and Chandra data was made for all sources with $\log(L_{\rm X}/L_{\rm BOL})>-7$ and $HR>0.2$ (a larger range of values, to allow for an in-depth check). Besides cases already published \citep{naz18,naz20,naz22}, Chandra data of CD--29\,5159 as well as XMM data of HD\,302798, V870\,Cen, $\mu^{02}$\,Cru, and V955\,Cen  were found in the archives. For V870\,Cen, the detected XMM source appears at 5\arcsec\ of the Be star and since it lies in the IC\,2944 cluster, the association may be spurious. Indeed, another 2MASS source is proposed to be associated with that source (see \citealt{naz13}). For the four other cases, the spectra were extracted in a standard way (for Chandra) or downloaded (pipeline products of XMM). They were then grouped to get at least 10 counts per bin and examined within {\sc Xspec} as the eROSITA spectra. The results are provided at the bottom of Table \ref{specfit}. The properties of CD--29\,5159 derived by both spacecrafts fully agree. HD\,302798 appears at least twice brighter in XMM data than in eROSITA survey. It is also slightly harder. In fact, at the time of XMM observations, this star fulfills all "bright \& hard" criteria while at the time of eROSITA, its $HR$ is just below the limit (but compatible with it within errors). Both $\mu^{02}$\,Cru and V955\, Cen appear slightly softer at the time of XMM observations compared to the time of eROSITA observations, as well as fainter, with a luminosity reduced by one third. They remain in their respective categories, however. The spectral fitting of CD--29 5159, HD\,302798, and V955\,Cen also clearly confirms the presence of an intense X-ray emission at high energies. This therefore validates the chosen criteria to identify bright and hard sources.

\begin{figure}
\includegraphics[width=9cm]{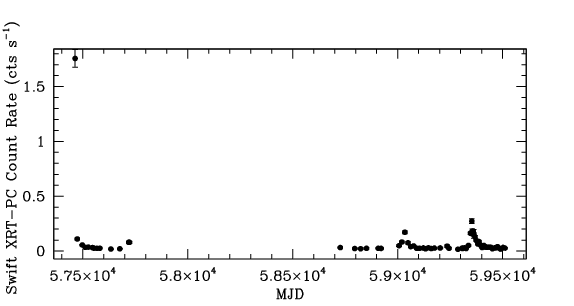}
\caption{Light curve of CD--29\,5159 recorded by the PC camera of the XRT onboard the Swift satellite. Note the several outbursts.}
\label{lccd}
\end{figure}

In addition, we also checked Swift archives although only for sources with $V<8.3$ to avoid optical loading in Swift X-ray telescope (XRT). Six sources were found in the Second Swift-XRT Point Source Catalog (2SXPS, \citealt{eva20}). For ten additional objects, some archival observations are available and count rates were derived using the on-line Swift product building tool\footnote{https://www.swift.ac.uk/user\_objects/} in a standard way. In both cases, the average count rates were converted into fluxes using WebPIMMS\footnote{https://heasarc.gsfc.nasa.gov/cgi-bin/Tools/w3pimms/w3pimms.pl} in the same way as done in 2SXPS, i.e. using interstellar absorption and either a power law of $\gamma=1.7$ or an $apec$ with $kT=1$\,keV. Table \ref{swift} provides the results. Comparing the eROSITA and Swift observed fluxes in 0.5--5.0\,keV, a large variation is detected for HD\,305560 and mild changes (2--3\,$\sigma$) for HD\,251726, CD--29\,5159, WRAY\,15--162, HD\,86689, V1083\,Sco, and HD\,316341. With few Swift counts in general, it is rather difficult to compare hardness, however. Only two objects have enough counts to derive meaningful spectra: CD--29\,5159 and HD\,305560. The former star appears extremely variable (Fig. \ref{lccd}) hence the average spectrum was not used. For the latter star, the average spectrum was built using the online tool and fitting results are provided in Table \ref{specfit}. This confirmed the huge increase in flux (2.6\,dex) already hinted by the count rate and also revealed a parallel increase of 50\% in $HR$. 

\begin{table*}
  \scriptsize
  \caption{Swift properties of some targets in common.  \label{swift}}
  \begin{tabular}{lccccc}
    \hline
Name  & $N^{\rm ISM}_{\rm H}$ & 2SXP Name & XRT-PC count rate & \multicolumn{2}{c}{$F_{\rm X}^{obs}$(erg\,cm$^{-2}$\,s$^{-1}$)}  \\
     &($10^{22}$\,cm$^{-2}$) & & ($10^{-2}$cts\,s$^{-1}$) & pow & apec \\
\hline
V1167\,Tau	&0.22&	2SXPS J055330.8+254434	&1.23$\pm$0.34 & (3.63$\pm$1.00)e-13& (2.62$\pm$0.72)e-13 \\
HD\,251726	&0.44&		                &1.16$\pm$0.48 & (3.65$\pm$1.51)e-13& (2.55$\pm$1.06)e-13 \\
CD--23\,6121	&0.30&		                &0.53$\pm$0.40 & (1.62$\pm$1.22)e-13& (1.15$\pm$0.87)e-13 \\
CD--29\,5159	&0.48&	2SXPS J075542.5--293353	&7.70$\pm$0.15 & (2.44$\pm$0.05)e-12& (1.70$\pm$0.03)e-12 \\
WRAY\,15--162	&0.73&		                &1.03$\pm$0.26 & (3.41$\pm$0.86)e-13& (2.37$\pm$0.60)e-13 \\
HD\,69425	&0.14&		                &0.23$\pm$0.17 & (6.46$\pm$4.78)e-14& (4.77$\pm$3.53)e-14 \\
QR\,Vel	        &0.17&	2SXPS J092158.3--511035	&0.64$\pm$0.23 & (1.85$\pm$0.66)e-13& (1.35$\pm$0.48)e-13 \\
HD\,86689	&0.23&		                &1.52$\pm$0.60 & (4.50$\pm$1.78)e-13& (3.25$\pm$1.28)e-13 \\
HD\,300584	&0.48&		                &1.35$\pm$0.59 & (4.28$\pm$1.87)e-13& (2.99$\pm$1.31)e-13 \\
HD\,305560	&0.36&	2SXPS J104628.5--603350	& 374$\pm$6.81 & (1.15$\pm$0.02)e-10& (8.13$\pm$0.15)e-11 \\
HD\,303763	&0.22&		                &1.29$\pm$1.22 & (3.80$\pm$3.59)e-13& (2.75$\pm$2.60)e-13 \\
V955\,Cen	&0.23&	2SXPS J131412.4--632223	&1.75$\pm$0.47 & (5.19$\pm$1.38)e-13& (3.74$\pm$1.00)e-13 \\
HD\,122669	&0.25&		                &1.40$\pm$0.77 & (4.18$\pm$2.30)e-13& (3.00$\pm$1.65)e-13 \\
HD\,154154	&0.21&		                &1.01$\pm$0.31 & (2.97$\pm$0.91)e-13& (2.15$\pm$0.66)e-13 \\
V1083\,Sco	&0.38&		                &2.83$\pm$1.07 & (8.77$\pm$3.32)e-13& (6.17$\pm$2.33)e-13 \\
HD\,316341	&0.41&	2SXPS J174835.3--295731	&0.30$\pm$0.18 & (9.27$\pm$5.56)e-14& (6.51$\pm$3.90)e-14 \\
\hline
  \end{tabular}
  
{\scriptsize Swift count rates are provided in the 0.3--10.\,keV band. Observed fluxes in 0.5--5.\,keV are calculated as in 2SXPS, i.e. considering either a power law of $\gamma=1.7$ or an $apec$ with $kT=1$\,keV, both absorbed by the known interstellar absorption.}
\end{table*}

Once the specific group of "bright \& hard" sources is identified, the following step is to constrain their nature. A major candidate is of course the \gc\ phenomenon, but we cannot directly exclude the presence of low-luminosity X-ray binaries. Fortunately, they display very different properties. The X-ray emission of the \gc\ objects is thermal in nature, as notably evidenced by the presence of the iron line near 6.7\,keV. The plasma temperatures generally lie in 5--20\,keV. Variability is ubiquitous in these stars, with flux generally changing by a factor of a few on different timescales (short-term "shots" down to a few seconds, long-term slow changes over months or years). The most extreme case is PZ\,Gem, in which a near-disappearance of the \gc\ characteristics was detected (which remains to present day, see above) as the Be disk disappeared. Regarding X-ray binaries, our initial catalog excluded the well known cases and no object is detected with X-ray luminosities above $10^{34}$\,erg\,s$^{-1}$. However, that does not prevent some unknown or poorly known X-ray binaries with (very) low X-ray luminosities (outside outbursts) to appear in our "bright \& hard" sample. Such objects should display a power law spectrum in the X-ray range and, over long timescales, exhibit large increases in flux (outbursts). 

While spectra and light curves are missing for most objects, the sample contains a few known cases. On the one hand, there are seven known \gc\ stars (see "\gc" in Name column of Table \ref{Targets}). Their spectral characteristics have been determined elsewhere (e.g. \citealt{naz18,naz20}): the best-fit temperatures remain within 7--13\,keV, except for V810\,Cas. At least hints of the iron line have been reported in several cases, including V810\,Cas. There is thus little doubt on their classification. On the other hand, a literature search also highlighted four other objects with previous analyses. CD--29\,5159 has been the focus of several recent studies \citep{dor21,ric23}. It is now considered as an X-ray binary with the companion, in a 59.5\,d orbit, being a pulsar exhibiting radio pulsations with $\sim$300\,s period. CD--29\,5159 also displays a highly variable light curve with clear outbursts (Fig. \ref{lccd}) and a very hard spectrum without trace of the iron line (the thermal fitting converges towards an extremely high temperature, see Table \ref{specfit}). A power law provides an equally good fitting to that Chandra spectrum, with an index $\Gamma=1.27\pm0.08$, quite typical of X-ray binaries. HD\,305560 was reported a few years ago to have been seen bursting in Swift data \citep{mas14a,mas14b}, although no follow-up was made. The fitting to the Swift spectra, which show no trace of the iron line, converges towards an extremely high temperature if a thermal fit is done (Table \ref{specfit}), while a power law fit yields $\Gamma=0.45\pm0.10$. As already mentioned, the flux at the time of the Swift exposures is much higher than during eROSITA observations: the source clearly was undergoing an outburst at the time. HD\,141926 is listed amongst HMXBs in \citet{liu06}, on the basis of a detection by the High Energy Astronomy Observatory (HEAO\,A-1) satellite \citep{woo84}. In that HEAO\,A-1 catalog, it appears under the name 1H1555--552 with a count rate of 0.0078$\pm$0.0004\,cts\,cm$^{-2}$\,s$^{-1}$, which would correspond to a flux of $3.7 \times 10^{-11}$\,erg\,m$^{-2}$\,s$^{-1}$ in the 2.--10.\,keV range if assuming a Crab-like spectrum. No follow-up was made since: a more thorough monitoring would be needed for a better assessment. Finally, HD\,316341 corresponds to the ASCA source AX\,J1748.6--2957, with a reported flux of $1.6 \times 10^{-12}$\,erg\,m$^{-2}$\,s$^{-1}$ in the 0.7--10.\,keV range and a best-fit power-law index $\Gamma=1.77$ \citep{sak02}. This flux appears one dex above the eROSITA value, suggesting that the source  was then observed during an outburst. However, it also appears as \#86 in \citet{sid01} with a count rate (0.1--2.4\,keV) of $0.0081\pm0.0014)$\,cts\,s$^{-1}$ in data from the R\"ontgen satellite - Position Sensitive Proportional Counter (ROSAT-PSPC) and as CXOGBS\,J174835.5--295728 in \citet{jon11} with 42 reported Chandra counts. It is also listed in the Chandra source catalog (CSC) v2.0 under the name 2CXO\,J174835.5--295729 with a flux in 0.5--7.\,keV of $(2.7\pm0.5) \times 10^{-13}$\,erg\,m$^{-2}$\,s$^{-1}$. That flux seems mostly located at high energies since the flux in 2--7.\,keV is evaluated at $(2.2\pm0.5) \times 10^{-13}$\,erg\,m$^{-2}$\,s$^{-1}$. In fact, the ROSAT, Chandra, and eROSITA detections present very similar fluxes, considering errors. Nevertheless, the ASCA X-ray source is also presented as one out of eighteen sources within the outbursting integral source H1743--322 \citep{par03}. The X-ray binary status of that system thus remains uncertain and additional observations should help clarifying the situation.

All in all, there are thus seven known \gc\ stars and possibly 2--4 low-luminosity X-ray binaries in our "bright \& hard" sample. This makes at most a proportion of two-thirds/one-third between these categories. If applied to the total number of 34 sources, that makes 23 \gc\ stars for 11 low-luminosity X-ray binaries, which would then triple the numbers in each category. To ascertain the conclusions, however, high-quality X-ray spectra and a long-term monitoring are required, as only such data can reveal the nature of the spectrum (power law/thermal) and the presence of outbursts. 

At the other extreme, one could wonder whether extremely soft sources are present in the sample. Indeed, studies in the Magellanic Clouds (MCs) detected a few supersoft sources associated to Be stars, the X-ray emission being linked to a WD companion \citep{kah06,stu12,coe20,ken21}. To this aim, a softness ratio $SR$ was calculated for the targets. It is calculated as the ratio between the flux in the 0.2--0.5\,keV and that in 0.5--5.\,keV (Table \ref{Targets}). For the literature supersoft sources mentioned above, a blackbody temperature of about 90\,eV provided a good fit and such a spectrum yields $SR$ well above unity. In our sample, 86\% of the targets display $SR<1$. Amongst the remaining sources, only five possess ratios significantly different (i.e. at least at two sigma) from the null value: $\kappa$\,CMa, 19\,Mon, $o$\,Pup, HD\,305483, and V1075\,Sco. Estimating the softness of the X-ray emission presents however two difficulties. On the one hand, optical loading by optically-bright sources will lead to an increase of the recorded X-rays at low energies. This effect has been corrected in our data but the correction remains far from perfect. This may cast doubt on the softness detection of the optically bright stars $\kappa$\,CMa, 19\,Mon, $o$\,Pup, and V1075\,Sco ($V=3.9-5.5$\,mag). On the other hand, fluxes were also corrected for the interstellar absorption. The fluxes at the lowest energies are particularly sensitive to this correction and larger absorptions will lead to larger uncertainties. In this context, it should be noted that two of the five cases (HD\,305483 and V1075\,Sco) display absorbing columns of $\sim2\times10^{21}$\,cm$^{-2}$. Finally, the MCs supersoft sources mentioned above all displayed very bright X-ray emissions ($L_{\rm X}$ of $10^{34}-10^{37}$\,erg\,s$^{-1}$) while all our sources with $SR>1$, including the five "best cases" just mentioned, have $L_{\rm X}<10^{32}$\,erg\,s$^{-1}$. There is thus little support for the presence of bright supersoft X-ray emitters, similar to those detected in the MCs, in association to any of our targets. 

\section{Summary and conclusions}
The eROSITA catalog compiling four semesters of sky survey observations was investigated to find counterparts of Be stars. To this aim, we used the list of Galactic Be stars from the BeSS database: 832 stars lie in the eROSITA\_DE half-sky, and 170 of them were detected in the X-ray range by eROSITA (detection fraction of about 20\%). Early-type stars and/or nearby objects have a higher chance to be detected, as could be expected. 

For all of these detected X-ray sources, the X-ray luminosities in the 0.5--5.\,keV energy range and corrected for interstellar absorption, the X-ray to bolometric luminosity ratios, and the hardness ratios (defined as the fraction of the 0.5--5.\,keV flux lying in the 2.--5.\,keV range) were estimated. The derived values cover a similar range as found for previous Be samples. Several checks were performed, comparing values derived from spectral analysis to those derived from counts, or comparing published values with eROSITA parameters for the 27 sources in common. All revealed a good agreement, considering the different upper limit (5/10\,keV), the errors and some (limited) stellar variability.

No obvious clear-cut is apparent when comparing X-ray luminosities to X-ray to bolometric luminosity ratios. There is only a clear correlation between them, but no obvious "grouping". On the contrary, hardness ratios reveal a splitting between a group of very soft (and fainter on average) sources and a smaller group of hard (and brighter on average) sources. The former group gathers most of the X-ray sources, but 34 sources can certainly be considered bright and hard. Amongst them are 7 known \gc\ stars and 2--4 low-luminosity X-ray binaries: the new census therefore triple the number of bright and hard objects, demonstrating that hard X-rays associated to Be stars are no so rare. Those 34 sources mostly correspond to Be stars with early spectral types, in line with e.g. the previously reported prevalence of the \gc\ phenomenon at higher stellar temperatures. Focusing on the most complete sample (stars with distances of 100--1000\,pc), an incidence rate of $\sim$12\% is derived for the bright and hard X-ray emissions associated to early-type Be stars. 

Finally, we have investigated the fluxes recorded at very low energies (0.2--0.5\,keV). Only a few targets are found to emit in this range, but the high uncertainties associated to these very soft fluxes make many of these detections insignificant. Furthermore, uncertainties linked to the optical loading correction and to the interstellar absorption correction eliminate the few cases of apparently soft emissions. Besides, all of these "very soft" candidates display luminosities below $10^{32}$\,erg\,s$^{-1}$. Therefore, our sample does not contain any obvious case of bright supersoft X-ray emission such as had been detected in the Magellanic Clouds and which were thought to be associated to Be+WD systems.

\section*{Acknowledgements}
This work is based on data from eROSITA, the soft X-ray instrument aboard SRG, a joint Russian-German science mission supported by the Russian Space Agency (Roskosmos), in the interests of the Russian Academy of Sciences represented by its Space Research Institute (IKI), and the Deutsches Zentrum f{\"u}r Luft- und Raumfahrt (DLR). The SRG spacecraft was built by Lavochkin Association (NPOL) and its subcontractors, and is operated by NPOL with support from the Max Planck Institute for Extraterrestrial Physics (MPE).
The development and construction of the eROSITA X-ray instrument was led by MPE, with contributions from the Dr. Karl Remeis Observatory Bamberg \& ECAP (FAU Erlangen-Nuernberg), the University of Hamburg Observatory, the Leibniz Institute for Astrophysics Potsdam (AIP), and the Institute for Astronomy and Astrophysics of the University of Tuebingen, with the support of DLR and the Max Planck Society. The Argelander Institute for Astronomy of the University of Bonn and the Ludwig Maximilians Universitaet Munich also participated in the science preparation for eROSITA. The eROSITA data shown here were processed using the eSASS software system developed by the German eROSITA consortium.

Y.N. acknowledges support from the Fonds National de la Recherche Scientifique (Belgium), the European Space Agency (ESA) and the Belgian Federal Science Policy Office (BELSPO) in the framework of the PRODEX Programme (contracts linked to XMM-Newton). J.R. acknowledges support from the DLR under grant 50QR2105.
ADS and CDS were used for preparing this document. This work has made use of the BeSS database, operated at LESIA, Observatoire de Meudon, France (http://basebe.obspm.fr).

\section*{Data availability}
The eROSITA data used in this article are available upon reasonable request. XMM, Chandra, and Swift data are publicly available in their respective archives.

\setcounter{table}{0}
\begin{landscape}
\begin{table}
  \scriptsize
  \caption{Target properties, ranked by increasing right ascension. } \label{Targets}
  \begin{tabular}{lccccccccccccc}
    \hline
 Name & SpT & flags & $d$ (pc) & $\log(L_{\rm BOL}$ & $N^{\rm ISM}_{\rm H}$ & \multicolumn{5}{c}{Rate (counts\,s$^{-1}$)} & \multicolumn{3}{c}{$F^{ISM\,cor}_{\rm X}$ (erg\,cm$^{-2}$\,s$^{-1}$)} \\  
      &     &       &          &  $/L_{\odot})$           & ($10^{22}$ & (0.2--2.3) & (0.2--0.5) & (0.5--1.) & (1.--2.) & (2.--5.) & (0.2--0.5) & (0.5--2.) & (0.5--5.) \\
      &     &       &          &                               & cm$^{-2}$)  \\      
\hline
 HD\,19818 & B9.5Vne & 0\_0\_0		       & 301.1$\pm$2.0   & 1.394$\pm$0.006 & 0.010 &  0.440$\pm$0.021 & 0.059$\pm$0.008 & 0.227$\pm$0.015 &  0.148$\pm$0.012 &  0.007$\pm$0.004 & (8.89$\pm$1.19)e-14 & (2.93$\pm$0.15)e-13 & (3.16$\pm$0.16)e-13 \\ 
 228\,Eri & B2Vne & 2\_0\_0		       & 471.7$\pm$16.7  & 3.953$\pm$0.031 & 0.031 &  0.013$\pm$0.006 & 0.007$\pm$0.004 & 0.003$\pm$0.003 &  0.001$\pm$0.002 &  0.000$\pm$0.004 & (1.56$\pm$0.85)e-14 & (3.11$\pm$2.73)e-15 & (3.11$\pm$4.19)e-15 \\ 
 $\lambda$\,Eri & B2III(e)p & 0\_0\_1	       & 248.8$\pm$11.1  & 3.774$\pm$0.039 & 0.013 &  0.109$\pm$0.015 &                 & 0.045$\pm$0.010 &  0.022$\pm$0.007 &  0.010$\pm$0.007 &                     & (5.02$\pm$0.90)e-14 & (1.18$\pm$0.21)e-13 \\ 
 AN\,Col & B2Vnpe & 0\_0\_0		       & 371.2$\pm$9.5   & 3.419$\pm$0.022 & 0.0037&  0.032$\pm$0.007 & 0.006$\pm$0.003 & 0.021$\pm$0.006 &  0.003$\pm$0.003 &  0.004$\pm$0.004 & (8.13$\pm$4.10)e-15 & (1.91$\pm$0.49)e-14 & (1.96$\pm$0.51)e-14 \\ 
 V1369\,Ori & B7Ib/II & 1\_0\_0		       & 419.8$\pm$7.4   & 2.971$\pm$0.015 & 0.036 &  0.017$\pm$0.007 & 0.002$\pm$0.002 & 0.006$\pm$0.004 &  0.011$\pm$0.006 &  0.000$\pm$0.002 & (4.22$\pm$5.07)e-15 & (1.42$\pm$0.60)e-14 & (1.45$\pm$0.64)e-14 \\ 
 120\,Tau & B2IVe & 1\_0\_0		       & 389.6$\pm$14.0  & 3.772$\pm$0.031 & 0.078 &  0.015$\pm$0.007 & 0.016$\pm$0.009 & 0.010$\pm$0.005 &  0.000$\pm$0.002 &  0.000$\pm$0.002 & (6.36$\pm$3.44)e-14 & (9.09$\pm$5.30)e-15 & (9.33$\pm$5.73)e-15 \\ 
 43\,Ori & O9.5IVp & 0\_2\_0		       & 473.9$\pm$92.1  & 3.966$\pm$0.169 & 0.035 &  1.142$\pm$0.234 & 0.046$\pm$0.010 & 0.522$\pm$0.070 &  0.299$\pm$0.031 &  0.033$\pm$0.013 & (1.07$\pm$0.24)e-13 & (6.47$\pm$0.61)e-13 & (7.38$\pm$0.68)e-13 \\ 
 HD\,245310 & B2IIInnpe & 0\_0\_0	       &1794.9$\pm$61.5  & 4.047$\pm$0.030 & 0.23  &  0.112$\pm$0.018 &                 & 0.019$\pm$0.008 &  0.085$\pm$0.016 &  0.027$\pm$0.011 &                     & (1.34$\pm$0.23)e-13 & (2.66$\pm$0.42)e-13 \\
 $\zeta$\,Tau (\gc) & B1IVe shell & 0\_0\_2    & 136.4$\pm$15.3  & 4.024$\pm$0.097 & 0.019 &  0.730$\pm$0.041 &                 & 0.000$\pm$0.011 &  0.028$\pm$0.010 &  0.284$\pm$0.031 &                     & (2.92$\pm$1.01)e-14 & (1.78$\pm$0.18)e-12 \\ 
 $\omega$\,Ori & B3Ve & 1\_0\_1		       & 497.3$\pm$43.3  & 4.243$\pm$0.076 & 0.086 &  0.031$\pm$0.009 &                 & 0.008$\pm$0.005 &  0.006$\pm$0.004 &  0.000$\pm$0.002 &                     & (1.21$\pm$0.53)e-14 & (1.24$\pm$0.58)e-14 \\ 
 HD\,39018 & B9 & 1\_0\_0		       & 332.7$\pm$5.5   & 2.081$\pm$0.014 & 0.028 &  0.020$\pm$0.008 & 0.002$\pm$0.003 & 0.010$\pm$0.005 &  0.005$\pm$0.004 &  0.003$\pm$0.005 & (4.41$\pm$5.62)e-15 & (1.25$\pm$0.54)e-14 & (1.29$\pm$0.56)e-14 \\ 
 V1167\,Tau & B1Vnne & 0\_0\_0		       &1089.2$\pm$49.0  & 4.049$\pm$0.039 & 0.22  &  0.147$\pm$0.020 & 0.002$\pm$0.003 & 0.022$\pm$0.008 &  0.116$\pm$0.018 &  0.041$\pm$0.013 & (3.08$\pm$3.32)e-14 & (1.71$\pm$0.24)e-13 & (4.07$\pm$0.52)e-13 \\
 HD\,250980 & B2ne & 1\_0\_0		       &2027.5$\pm$86.6  & 3.900$\pm$0.037 & 0.20  &  0.029$\pm$0.010 &                 & 0.003$\pm$0.003 &  0.022$\pm$0.009 &  0.006$\pm$0.006 &                     & (3.15$\pm$1.16)e-14 & (3.24$\pm$1.13)e-14 \\
 HD\,251726 & B1Ve & 1\_0\_0		       &2094.1$\pm$88.8  & 4.575$\pm$0.037 & 0.44  &  0.039$\pm$0.012 &                 & 0.007$\pm$0.006 &  0.025$\pm$0.009 &  0.013$\pm$0.009 &                     & (6.17$\pm$2.04)e-14 & (9.55$\pm$2.89)e-14 \\
 HD\,42054 (\gc\,cand) & B5Ve & 0\_0\_0	       & 291.0$\pm$50.9  & 3.036$\pm$0.152 & 0.0055&  1.291$\pm$0.045 & 0.144$\pm$0.015 & 0.563$\pm$0.030 &  0.545$\pm$0.030 &  0.094$\pm$0.014 & (1.95$\pm$0.21)e-13 & (8.66$\pm$0.33)e-13 & (1.37$\pm$0.05)e-12 \\ 
 HD\,253659 & B0.5Vnne & 1\_0\_0	       &1678.7$\pm$79.1  & 4.842$\pm$0.041 & 0.61  &  0.039$\pm$0.012 & 0.002$\pm$0.003 & 0.006$\pm$0.005 &  0.023$\pm$0.009 &  0.011$\pm$0.008 & (1.48$\pm$2.71)e-13 & (6.45$\pm$2.26)e-14 & (9.44$\pm$3.08)e-14 \\
 HD\,43285 & B6Ve & 0\_0\_0		       & 213.0$\pm$3.0   & 2.595$\pm$0.012 & 0.0092&  0.181$\pm$0.023 & 0.017$\pm$0.007 & 0.106$\pm$0.018 &  0.055$\pm$0.013 &  0.003$\pm$0.004 & (2.51$\pm$1.06)e-14 & (1.27$\pm$0.17)e-13 & (1.30$\pm$0.18)e-13 \\ 
 HD\,43544 & B2.5Ve & 2\_0\_0		       & 307.4$\pm$5.1   & 3.259$\pm$0.014 & 0.021 &  0.013$\pm$0.007 &                 & 0.010$\pm$0.005 &  0.003$\pm$0.003 &  0.000$\pm$0.002 &                     & (1.03$\pm$0.50)e-14 & (1.03$\pm$0.52)e-14 \\ 
 HD\,254647 & Bpe & 1\_0\_0		       &2437.9$\pm$102.7 & 4.191$\pm$0.037 & 0.36  &  0.059$\pm$0.014 & 0.002$\pm$0.003 & 0.004$\pm$0.004 &  0.050$\pm$0.013 &  0.024$\pm$0.010 & (7.74$\pm$15.4)e-14 & (8.26$\pm$2.06)e-14 & (2.30$\pm$0.50)e-13 \\
 HD\,44506 & B3V & 1\_0\_0	               & 552.4$\pm$27.3  & 3.822$\pm$0.043 & 0.017 &  0.018$\pm$0.006 & 0.007$\pm$0.004 & 0.010$\pm$0.005 &  0.000$\pm$0.002 &  0.003$\pm$0.005 & (1.18$\pm$0.67)e-14 & (6.22$\pm$2.90)e-15 & (6.33$\pm$3.06)e-15 \\ 
 FR\,CMa (\gc) & B1.5IVe & 0\_0\_0             & 533.6$\pm$16.1  & 4.217$\pm$0.026 & 0.054 &  1.139$\pm$0.053 & 0.060$\pm$0.012 & 0.364$\pm$0.030 &  0.643$\pm$0.040 &  0.178$\pm$0.025 & (1.74$\pm$0.36)e-13 & (9.19$\pm$0.46)e-13 & (1.76$\pm$0.08)e-12 \\ 
 BD--21$^{\circ}$1449 & Be & 1\_0\_0	       &2743.5$\pm$183.3 & 3.630$\pm$0.058 & 0.12  &  0.089$\pm$0.021 & 0.010$\pm$0.007 & 0.029$\pm$0.012 &  0.043$\pm$0.015 &  0.000$\pm$0.010 & (7.06$\pm$4.89)e-14 & (7.68$\pm$2.03)e-14 & (7.87$\pm$2.35)e-14 \\
 PZ\,Gem (\gc) & O9:npe & 0\_0\_0	       & 855.6$\pm$25.7  & 4.523$\pm$0.026 & 0.086 &  0.101$\pm$0.018 & 0.001$\pm$0.004 & 0.031$\pm$0.010 &  0.064$\pm$0.014 &  0.009$\pm$0.007 & (5.89$\pm$17.0)e-15 & (8.80$\pm$1.58)e-14 & (9.01$\pm$1.58)e-14 \\ 
 HD\,45995 (\gc) & B1.5Vne & 0\_0\_0	       & 655.1$\pm$21.8  & 4.169$\pm$0.029 & 0.059 &  0.375$\pm$0.034 & 0.021$\pm$0.008 & 0.146$\pm$0.021 &  0.200$\pm$0.025 &  0.046$\pm$0.014 & (6.49$\pm$2.54)e-14 & (3.08$\pm$0.29)e-13 & (5.80$\pm$0.53)e-13 \\ 
 QQ\,Gem & Be & 2\_0\_0			       & 555.6$\pm$11.4  & 3.090$\pm$0.018 & 0.017 &  0.021$\pm$0.014 &                 & 0.000$\pm$0.000 &  0.000$\pm$0.000 &  0.000$\pm$0.000 &                     & (1.71$\pm$1.16)e-14 & (1.79$\pm$1.22)e-14 \\ 
 HD\,46484 & B0.5IVe & 2\_0\_0		       &1366.1$\pm$53.0  & 4.646$\pm$0.034 & 0.22  &  0.013$\pm$0.007 &                 & 0.008$\pm$0.005 &  0.004$\pm$0.004 &  0.000$\pm$0.003 &                     & (1.60$\pm$0.84)e-14 & (1.60$\pm$0.90)e-14 \\
 HD\,47054 & B8IVe & 0\_0\_0		       & 226.6$\pm$5.0   & 2.725$\pm$0.019 & 0.017 &  0.075$\pm$0.015 & 0.018$\pm$0.007 & 0.044$\pm$0.011 &  0.011$\pm$0.006 &  0.003$\pm$0.005 & (3.00$\pm$1.25)e-14 & (4.09$\pm$0.97)e-14 & (4.19$\pm$1.01)e-14 \\ 
 V733\,Mon & B3(II)e & 0\_0\_0		       &2059.7$\pm$93.7  & 4.135$\pm$0.040 & 0.25  &  0.070$\pm$0.015 & 0.013$\pm$0.006 & 0.020$\pm$0.008 &  0.042$\pm$0.012 &  0.011$\pm$0.008 & (2.23$\pm$1.07)e-13 & (7.71$\pm$1.74)e-14 & (1.58$\pm$0.35)e-13 \\
 15\,Mon & O7V+B1.5/2V & 0\_0\_1	       & 281.7$\pm$39.7  & 4.301$\pm$0.122 & 0.0067&  3.768$\pm$0.107 & 0.817$\pm$0.049 & 2.154$\pm$0.146 &  0.557$\pm$0.043 &  0.032$\pm$0.014 & (1.14$\pm$0.07)e-12 & (1.96$\pm$0.11)e-12 & (2.08$\pm$0.12)e-12 \\ 
 10\,CMa & B2V & 1\_0\_0		       & 604.7$\pm$26.7  & 4.253$\pm$0.038 & 0.024 &  0.018$\pm$0.007 & 0.010$\pm$0.004 & 0.006$\pm$0.004 &  0.000$\pm$0.003 &  0.000$\pm$0.002 & (1.90$\pm$0.84)e-14 & (4.15$\pm$3.52)e-15 & (4.25$\pm$3.93)e-15 \\ 
 HP\,CMa & B1.5Vne & 1\_0\_0		       & 599.2$\pm$33.9  & 4.205$\pm$0.049 & 0.024 &  0.019$\pm$0.007 & 0.006$\pm$0.004 & 0.007$\pm$0.004 &  0.005$\pm$0.003 &  0.000$\pm$0.005 & (1.23$\pm$0.83)e-14 & (1.04$\pm$0.46)e-14 & (1.06$\pm$0.61)e-14 \\ 
 V715\,Mon & B2.5III & 1\_0\_0		       & 732.9$\pm$29.3  & 3.909$\pm$0.035 & 0.031 &  0.024$\pm$0.010 & 0.005$\pm$0.005 & 0.010$\pm$0.006 &  0.008$\pm$0.005 &  0.003$\pm$0.004 & (1.16$\pm$1.04)e-14 & (1.57$\pm$0.72)e-14 & (1.61$\pm$0.73)e-14 \\ 
 KS\,CMa & B9e & 0\_0\_0		       & 548.1$\pm$8.9   & 2.731$\pm$0.014 & 0.046 &  0.190$\pm$0.022 & 0.015$\pm$0.006 & 0.100$\pm$0.016 &  0.068$\pm$0.014 &  0.005$\pm$0.006 & (3.96$\pm$1.63)e-14 & (1.46$\pm$0.18)e-13 & (1.50$\pm$0.19)e-13 \\ 
 $\kappa$\,CMa & B1.5Ve & 0\_0\_2	       & 202.0$\pm$4.9   & 3.933$\pm$0.021 & 0.0031&  0.236$\pm$0.021 & 0.109$\pm$0.011 & 0.018$\pm$0.004 &  0.018$\pm$0.006 &  0.000$\pm$0.002 & (1.41$\pm$0.14)e-13 & (2.91$\pm$0.63)e-14 & (2.99$\pm$0.66)e-14 \\ 
 HZ\,CMa & B6IV)e+A & 0\_0\_0		       & 194.9$\pm$2.2   & 2.635$\pm$0.010 & 0.0031&  0.073$\pm$0.013 & 0.022$\pm$0.007 & 0.044$\pm$0.009 &  0.000$\pm$0.002 &  0.004$\pm$0.005 & (2.83$\pm$0.85)e-14 & (3.28$\pm$0.71)e-14 & (3.36$\pm$0.76)e-14 \\ 
 V742\,Mon & B2IIIe & 0\_0\_0		       & 977.2$\pm$24.7  & 4.073$\pm$0.022 & 0.097 &  0.545$\pm$0.041 & 0.013$\pm$0.007 & 0.204$\pm$0.025 &  0.297$\pm$0.030 &  0.066$\pm$0.017 & (7.61$\pm$4.31)e-14 & (4.95$\pm$0.39)e-13 & (8.18$\pm$0.62)e-13 \\ 
 V746\,Mon & B1V:nn & 1\_0\_0		       &1597.0$\pm$96.7  & 4.578$\pm$0.053 & 0.23  &  0.032$\pm$0.010 &                 & 0.014$\pm$0.007 &  0.017$\pm$0.007 &  0.001$\pm$0.003 &                     & (4.07$\pm$1.27)e-14 & (4.18$\pm$1.33)e-14 \\
 HD\,266894 & Be & 1\_0\_0	               &3702.9$\pm$391.2 & 3.959$\pm$0.092 & 0.19  &  0.021$\pm$0.009 &                 & 0.000$\pm$0.003 &  0.020$\pm$0.008 &  0.007$\pm$0.006 &                     & (2.59$\pm$1.13)e-14 & (2.67$\pm$1.04)e-14 \\
 HD\,51452 & B0IVe & 2\_0\_0		       &1924.1$\pm$137.0 & 4.941$\pm$0.062 & 0.25  &  0.015$\pm$0.007 &                 & 0.011$\pm$0.006 &  0.005$\pm$0.004 &  0.000$\pm$0.002 &                     & (2.24$\pm$1.05)e-14 & (2.24$\pm$1.10)e-14 \\
 19\,Mon & B2Vn(e) & 0\_0\_0		       & 370.1$\pm$19.3  & 3.884$\pm$0.045 & 0.020 &  0.089$\pm$0.016 & 0.032$\pm$0.010 & 0.042$\pm$0.011 &  0.013$\pm$0.006 &  0.000$\pm$0.003 & (5.72$\pm$1.71)e-14 & (4.19$\pm$0.98)e-14 & (4.29$\pm$1.03)e-14 \\ 
 27\,CMa & B4Ve sh & 0\_0\_1		       & 531.9$\pm$90.5  & 4.112$\pm$0.148 & 0.014 &  0.168$\pm$0.021 & 0.023$\pm$0.006 & 0.044$\pm$0.011 &  0.079$\pm$0.014 &  0.074$\pm$0.016 & (3.79$\pm$1.03)e-14 & (1.03$\pm$0.15)e-13 & (5.35$\pm$0.64)e-13 \\ 
 HD\,55806 & B7IIIe & 1\_0\_0		       & 856.5$\pm$16.1  & 2.572$\pm$0.016 & 0.039 &  0.040$\pm$0.012 & 0.007$\pm$0.005 & 0.015$\pm$0.007 &  0.017$\pm$0.008 &  0.003$\pm$0.005 & (1.61$\pm$1.21)e-14 & (2.85$\pm$0.99)e-14 & (2.92$\pm$1.01)e-14 \\ 
 NV\,Pup & B2V+B3IVne & 1\_0\_1		       & 246.3$\pm$10.9  & 3.634$\pm$0.039 & 0.0055&  0.033$\pm$0.009 &                 & 0.016$\pm$0.006 &  0.008$\pm$0.004 &  0.004$\pm$0.004 &                     & (2.03$\pm$0.64)e-14 & (2.09$\pm$0.64)e-14 \\ 
 HD\,57682 & O9.2IV & 0\_0\_0		       &1106.1$\pm$96.1  & 4.731$\pm$0.075 & 0.057 &  0.538$\pm$0.040 & 0.036$\pm$0.010 & 0.324$\pm$0.031 &  0.167$\pm$0.023 &  0.000$\pm$0.002 & (1.06$\pm$0.31)e-13 & (4.21$\pm$0.33)e-13 & (4.31$\pm$0.34)e-13 \\
 HD\,57910 & B5V & 1\_0\_0	               &1020.6$\pm$24.5  & 2.732$\pm$0.021 & 0.035 &  0.015$\pm$0.008 & 0.003$\pm$0.003 & 0.009$\pm$0.005 &  0.005$\pm$0.004 &  0.000$\pm$0.002 & (6.16$\pm$7.55)e-15 & (1.22$\pm$0.56)e-14 & (1.25$\pm$0.60)e-14 \\
 NO\,CMa & B3V & 1\_0\_0		       & 392.2$\pm$10.3  & 3.551$\pm$0.023 & 0.012 &  0.024$\pm$0.008 & 0.009$\pm$0.005 & 0.011$\pm$0.005 &  0.004$\pm$0.003 &  0.001$\pm$0.003 & (1.38$\pm$0.72)e-14 & (1.06$\pm$0.47)e-14 & (1.09$\pm$0.52)e-14 \\ 
 FY\,CMa & B1IIe & 0\_0\_0	               & 558.1$\pm$19.4  & 4.169$\pm$0.030 & 0.020 &  0.095$\pm$0.016 & 0.021$\pm$0.007 & 0.060$\pm$0.013 &  0.012$\pm$0.007 &  0.000$\pm$0.002 & (3.78$\pm$1.34)e-14 & (5.33$\pm$1.05)e-14 & (5.46$\pm$1.09)e-14 \\ 
 RY\,Gem & A0V:e+K0II & 0\_0\_0		       & 414.2$\pm$4.1   & 1.767$\pm$0.009 & 0.012 &  0.080$\pm$0.017 & 0.013$\pm$0.007 & 0.048$\pm$0.013 &  0.019$\pm$0.009 &  0.008$\pm$0.007 & (1.99$\pm$1.12)e-14 & (5.34$\pm$1.26)e-14 & (5.47$\pm$1.26)e-14 \\ 
 V378\,Pup & B2Ve & 2\_0\_0		       & 471.0$\pm$15.2  & 3.886$\pm$0.028 & 0.055 &  0.018$\pm$0.008 & 0.003$\pm$0.004 & 0.009$\pm$0.005 &  0.005$\pm$0.005 &  0.006$\pm$0.007 & (8.21$\pm$11.4)e-15 & (1.32$\pm$0.69)e-14 & (1.35$\pm$0.67)e-14 \\ 
 BN\,Gem & O8Vpev & 0\_0\_0		       &1787.7$\pm$157.1 & 5.091$\pm$0.076 & 0.070 &  0.063$\pm$0.017 & 0.019$\pm$0.009 & 0.026$\pm$0.010 &  0.014$\pm$0.007 &  0.009$\pm$0.007 & (6.73$\pm$3.10)e-14 & (3.76$\pm$1.18)e-14 & (3.86$\pm$1.15)e-14 \\
 HD\,62367 & B8V & 1\_0\_0	               & 427.4$\pm$7.2   & 2.636$\pm$0.015 & 0.019 &  0.023$\pm$0.009 &                 & 0.014$\pm$0.007 &  0.010$\pm$0.006 &  0.000$\pm$0.003 &                     & (1.93$\pm$0.74)e-14 & (1.98$\pm$0.80)e-14 \\ 
 HD\,62780 & O9/B0e & 1\_0\_0		       &2963.5$\pm$124.5 & 4.996$\pm$0.036 & 0.30  &  0.033$\pm$0.011 & 0.001$\pm$0.003 & 0.004$\pm$0.004 &  0.025$\pm$0.009 &  0.014$\pm$0.009 & (6.51$\pm$13.9)e-14 & (4.22$\pm$1.42)e-14 & (1.51$\pm$0.48)e-13 \\
 CD--23\,6121 & Be & 0\_0\_0		       &4100.0$\pm$281.9 & 4.387$\pm$0.060 & 0.30  &  0.060$\pm$0.014 &                 & 0.007$\pm$0.005 &  0.044$\pm$0.012 &  0.015$\pm$0.009 &                     & (8.22$\pm$2.06)e-14 & (1.22$\pm$0.29)e-13 \\
 V392\,Pup & B5V & 0\_0\_0	               & 181.2$\pm$7.9   & 2.610$\pm$0.038 & 0.0049&  0.045$\pm$0.010 & 0.003$\pm$0.003 & 0.034$\pm$0.008 &  0.006$\pm$0.004 &  0.005$\pm$0.005 & (3.91$\pm$3.78)e-15 & (2.88$\pm$0.66)e-14 & (2.95$\pm$0.68)e-14 \\ 
 $o$\,Pup & B1IVe & 0\_0\_1   	               & 434.8$\pm$43.5  & 4.438$\pm$0.087 & 0.015 &  0.087$\pm$0.017 & 0.027$\pm$0.007 & 0.039$\pm$0.011 &  0.000$\pm$0.002 &  0.000$\pm$0.004 & (4.35$\pm$1.13)e-14 & (2.85$\pm$0.82)e-14 & (2.92$\pm$0.90)e-14 \\ 
 CD--29\,5159 & B0Ve & 0\_0\_0		       &3289.4$\pm$128.1 & 5.007$\pm$0.034 & 0.48  &  0.222$\pm$0.024 &                 & 0.034$\pm$0.010 &  0.174$\pm$0.021 &  0.105$\pm$0.020 &                     & (3.68$\pm$0.40)e-13 & (9.95$\pm$0.96)e-13 \\
 V374\,Car & B3Vn & 0\_0\_0		       & 402.3$\pm$7.6   & 3.524$\pm$0.016 & 0.062 &  0.091$\pm$0.009 & 0.007$\pm$0.002 & 0.033$\pm$0.005 &  0.044$\pm$0.006 &  0.006$\pm$0.003 & (2.21$\pm$0.78)e-14 & (7.12$\pm$0.76)e-14 & (8.64$\pm$0.91)e-14 \\ 
 V408\,Pup & B2.5IIIe & 1\_0\_0		       &2650.3$\pm$170.6 & 3.981$\pm$0.056 & 0.13  &  0.030$\pm$0.010 & 0.002$\pm$0.003 & 0.009$\pm$0.005 &  0.020$\pm$0.008 &  0.003$\pm$0.004 & (1.75$\pm$1.96)e-14 & (2.99$\pm$1.02)e-14 & (3.06$\pm$1.01)e-14 \\
\hline
  \end{tabular}
\end{table}
\end{landscape}

\setcounter{table}{0}
\begin{landscape}
\begin{table}
  \scriptsize
  \caption{Continued. } 
  \begin{tabular}{lccccccccccccc}
    \hline
 Name & SpT & flags & $d$ (pc) & $\log(L_{\rm BOL}$ & $N^{\rm ISM}_{\rm H}$ & \multicolumn{5}{c}{Rate (counts\,s$^{-1}$)} & \multicolumn{3}{c}{$F^{ISM\,cor}_{\rm X}$ (erg\,cm$^{-2}$\,s$^{-1}$)} \\  
      &     &       &          &  $/L_{\odot})$           & ($10^{22}$ & (0.2--2.3) & (0.2--0.5) & (0.5--1.) & (1.--2.) & (2.--5.) & (0.2--0.5) & (0.5--2.) & (0.5--5.) \\
      &     &       &          &                               & cm$^{-2}$)  \\      
\hline
 CD--28\,5235 & B1V & 1\_0\_0		       &2714.8$\pm$105.5 & 4.207$\pm$0.034 & 0.27  &  0.032$\pm$0.010 &                 &  0.014$\pm$0.007 &  0.017$\pm$0.007 &  0.002$\pm$0.005 &                     & (4.28$\pm$1.36)e-14 & (4.39$\pm$1.44)e-14 \\
 CD--31\,5475 & B3Vep & 2\_0\_0		       &1470.2$\pm$29.4  & 3.136$\pm$0.017 & 0.081 &  0.020$\pm$0.008 & 0.002$\pm$0.002 &  0.010$\pm$0.006 &  0.009$\pm$0.005 &  0.004$\pm$0.005 & (8.32$\pm$9.94)e-15 & (1.82$\pm$0.72)e-14 & (1.82$\pm$0.69)e-14 \\
 WRAY\,15-162 & Bpe & 2\_0\_0		       &3155.2$\pm$90.7  & 4.195$\pm$0.025 & 0.73  &  0.013$\pm$0.006 &                 &  0.000$\pm$0.000 &  0.000$\pm$0.000 &  0.000$\pm$0.000 &                     & (3.15$\pm$1.56)e-14 & (1.70$\pm$0.84)e-13 \\
 V420\,Pup & B3Ve & 0\_0\_0		       & 501.8$\pm$9.2   & 3.374$\pm$0.016 & 0.036 &  0.530$\pm$0.038 & 0.051$\pm$0.012 &  0.217$\pm$0.024 &  0.248$\pm$0.026 &  0.039$\pm$0.013 & (1.20$\pm$0.28)e-13 & (3.94$\pm$0.30)e-13 & (6.08$\pm$0.45)e-13 \\ 
 Cl*\,Ruprecht\,55\,DE\,32 & B0IIIe & 1\_0\_0  &6308.9$\pm$1060.3& 5.000$\pm$0.146 & 0.40  &  0.032$\pm$0.010 &                 &  0.005$\pm$0.004 &  0.023$\pm$0.008 &  0.015$\pm$0.009 &                     & (4.91$\pm$1.53)e-14 & (1.33$\pm$0.38)e-13 \\
 HD\,69026 & B1.5Ve & 1\_0\_0		       & 887.0$\pm$16.4  & 3.698$\pm$0.016 & 0.15  &  0.016$\pm$0.005 & 0.001$\pm$0.002 &  0.001$\pm$0.002 &  0.009$\pm$0.004 &  0.004$\pm$0.004 & (7.70$\pm$12.9)e-15 & (1.39$\pm$0.55)e-14 & (2.47$\pm$0.90)e-14 \\ 
 HD\,69425 & B1Vpe & 1\_0\_0		       &1591.2$\pm$36.4  & 3.976$\pm$0.020 & 0.14  &  0.023$\pm$0.009 &                 &  0.016$\pm$0.007 &  0.007$\pm$0.004 &  0.002$\pm$0.004 &                     & (2.38$\pm$0.87)e-14 & (5.78$\pm$2.23)e-14 \\
 HD\,71072 & B4IIIe & 0\_0\_0		       & 587.9$\pm$14.0  & 3.288$\pm$0.021 & 0.024 &  0.032$\pm$0.011 &                 &  0.012$\pm$0.007 &  0.022$\pm$0.009 &  0.000$\pm$0.003 &                     & (2.92$\pm$0.95)e-14 & (3.00$\pm$1.00)e-14 \\ 
 HD\,71510 & B3IV & 1\_0\_0		       & 232.0$\pm$4.4   & 3.208$\pm$0.017 & 0.017 &  0.026$\pm$0.007 & 0.014$\pm$0.005 &  0.012$\pm$0.005 &  0.000$\pm$0.001 &  0.000$\pm$0.003 & (2.30$\pm$0.80)e-14 & (8.42$\pm$3.38)e-15 & (8.62$\pm$3.94)e-15 \\ 
 HD\,72063 & B2Vne & 0\_0\_0		       &1852.0$\pm$50.3  & 3.695$\pm$0.024 & 0.097 &  0.047$\pm$0.011 &                 &  0.019$\pm$0.007 &  0.028$\pm$0.008 &  0.005$\pm$0.005 &                     & (4.49$\pm$1.07)e-14 & (4.60$\pm$1.09)e-14 \\
 V471\,Car & B5ne & 0\_0\_0		       & 897.2$\pm$22.4  & 3.254$\pm$0.022 & 0.069 &  0.088$\pm$0.008 & 0.007$\pm$0.002 &  0.035$\pm$0.005 &  0.044$\pm$0.006 &  0.006$\pm$0.003 & (2.38$\pm$0.81)e-14 & (7.26$\pm$0.73)e-14 & (9.94$\pm$0.98)e-14 \\ 
 HD\,74401 & B1IIIne & 0\_0\_0		       &1767.3$\pm$41.5  & 3.977$\pm$0.020 & 0.13  &  0.151$\pm$0.021 & 0.012$\pm$0.008 &  0.042$\pm$0.013 &  0.064$\pm$0.012 &  0.035$\pm$0.010 & (8.73$\pm$5.95)e-14 & (1.14$\pm$0.19)e-13 & (2.96$\pm$0.42)e-13 \\
 HD\,74559 & B9Ve & 1\_0\_0		       & 693.8$\pm$8.9   & 2.215$\pm$0.011 & 0.058 &  0.054$\pm$0.014 &                 &  0.042$\pm$0.011 &  0.016$\pm$0.007 &  0.000$\pm$0.001 &                     & (4.17$\pm$0.94)e-14 & (4.25$\pm$0.97)e-14 \\ 
 HD\,75081 & B9V & 0\_0\_0	               & 220.5$\pm$1.7   & 2.292$\pm$0.007 & 0.0092&  0.110$\pm$0.016 & 0.016$\pm$0.006 &  0.062$\pm$0.011 &  0.027$\pm$0.008 &  0.012$\pm$0.008 & (2.42$\pm$0.95)e-14 & (7.03$\pm$1.08)e-14 & (1.38$\pm$0.21)e-13 \\ 
 HD\,75661 & B2Vne & 2\_0\_0		       &1762.3$\pm$53.4  & 3.672$\pm$0.026 & 0.13  &  0.009$\pm$0.004 &                 &  0.005$\pm$0.003 &  0.005$\pm$0.003 &  0.002$\pm$0.003 &                     & (1.10$\pm$0.48)e-14 & (1.10$\pm$0.49)e-14 \\
 HD\,75925 & B4Vnne & 2\_0\_0		       & 926.5$\pm$15.3  & 3.177$\pm$0.014 & 0.16  &  0.031$\pm$0.010 &                 &  0.021$\pm$0.008 &  0.015$\pm$0.007 &  0.000$\pm$0.004 &                     & (3.97$\pm$1.12)e-14 & (3.97$\pm$1.20)e-14 \\ 
 CD--45\,4826 & O9 & 1\_0\_0		       &2139.6$\pm$53.6  & 4.060$\pm$0.022 & 0.20  &  0.018$\pm$0.007 & 0.002$\pm$0.003 &  0.002$\pm$0.003 &  0.010$\pm$0.005 &  0.005$\pm$0.004 & (1.87$\pm$2.70)e-14 & (1.58$\pm$0.70)e-14 & (1.62$\pm$0.63)e-14 \\
 E\,Car & B3III & 0\_0\_1		       & 406.3$\pm$14.3  & 3.940$\pm$0.031 & 0.043 &  0.043$\pm$0.005 &                 &  0.023$\pm$0.004 &  0.009$\pm$0.003 &  0.004$\pm$0.003 &                     & (2.63$\pm$0.39)e-14 & (5.72$\pm$0.85)e-14 \\ 
 HD\,79066 & A9IV & 0\_0\_0		       &  50.7$\pm$0.1   & 0.762$\pm$0.001 & 0.0006&  0.204$\pm$0.028 & 0.055$\pm$0.014 &  0.130$\pm$0.022 &  0.017$\pm$0.008 &  0.000$\pm$0.003 & (6.64$\pm$1.72)e-14 & (1.03$\pm$0.16)e-13 & (1.06$\pm$0.17)e-13 \\ 
 HD\,79778 & B2Vne & 0\_0\_0		       &1586.8$\pm$50.6  & 3.679$\pm$0.028 & 0.13  &  0.139$\pm$0.014 & 0.008$\pm$0.004 &  0.048$\pm$0.008 &  0.076$\pm$0.010 &  0.027$\pm$0.008 & (5.38$\pm$2.61)e-14 & (1.28$\pm$0.13)e-13 & (2.79$\pm$0.28)e-13 \\
 HD\,80156 & B8.5IVe & 0\_0\_0		       & 499.5$\pm$6.1   & 2.083$\pm$0.011 & 0.023 &  0.049$\pm$0.012 &                 &  0.041$\pm$0.011 &  0.010$\pm$0.005 &  0.003$\pm$0.004 &                     & (4.14$\pm$0.95)e-14 & (4.25$\pm$0.98)e-14 \\ 
 QQ\,Vel & B2:nep & 1\_0\_0		       &4240.3$\pm$218.9 & 4.417$\pm$0.045 & 0.19  &  0.042$\pm$0.012 & 0.001$\pm$0.003 &  0.009$\pm$0.006 &  0.026$\pm$0.009 &  0.012$\pm$0.007 & (1.51$\pm$2.80)e-14 & (4.04$\pm$1.19)e-14 & (6.94$\pm$1.85)e-14 \\
 QR\,Vel & B2Vne & 0\_0\_0	               &1956.2$\pm$58.9  & 3.633$\pm$0.026 & 0.17  &  0.035$\pm$0.008 &                 &  0.006$\pm$0.004 &  0.026$\pm$0.007 &  0.007$\pm$0.005 &                     & (3.69$\pm$0.88)e-14 & (8.02$\pm$1.87)e-14 \\
 HD\,83060 & B2Vnne & 0\_0\_0		       &1354.5$\pm$40.3  & 3.562$\pm$0.026 & 0.11  &  0.056$\pm$0.009 &                 &  0.015$\pm$0.005 &  0.030$\pm$0.006 &  0.022$\pm$0.006 &                     & (5.08$\pm$0.89)e-14 & (1.48$\pm$0.23)e-13 \\
 I\,Hya & B5V & 0\_0\_1			       & 168.3$\pm$5.6   & 3.002$\pm$0.029 & 0.011 &  0.491$\pm$0.039 & 0.078$\pm$0.015 &  0.226$\pm$0.026 &  0.167$\pm$0.023 &  0.014$\pm$0.009 & (1.19$\pm$0.22)e-13 & (3.02$\pm$0.27)e-13 & (3.40$\pm$0.30)e-13 \\ 
 HD\,84567 & B0IV & 0\_0\_0		       &1070.9$\pm$55.0  & 4.660$\pm$0.045 & 0.043 &  0.048$\pm$0.012 & 0.008$\pm$0.005 &  0.035$\pm$0.010 &  0.004$\pm$0.004 &  0.000$\pm$0.005 & (2.13$\pm$1.30)e-14 & (2.99$\pm$0.83)e-14 & (3.06$\pm$0.93)e-14 \\
 HD\,85860 & B4Vne & 1\_0\_0		       & 460.3$\pm$8.2   & 3.011$\pm$0.015 & 0.030 &  0.026$\pm$0.010 & 0.003$\pm$0.005 &  0.011$\pm$0.006 &  0.012$\pm$0.007 &  0.000$\pm$0.002 & (7.12$\pm$10.2)e-15 & (2.07$\pm$0.81)e-14 & (2.13$\pm$0.85)e-14 \\ 
 HD\,86689 & B0/2ne & 0\_0\_0		       &2387.1$\pm$146.5 & 4.044$\pm$0.053 & 0.23  &  0.071$\pm$0.010 & 0.003$\pm$0.002 &  0.009$\pm$0.004 &  0.051$\pm$0.008 &  0.013$\pm$0.006 & (4.28$\pm$3.18)e-14 & (7.87$\pm$1.21)e-14 & (1.34$\pm$0.20)e-13 \\
 OY Hya & B5Ve & 1\_0\_0		       & 304.6$\pm$9.5   & 2.933$\pm$0.027 & 0.017 &  0.075$\pm$0.016 &                 &  0.060$\pm$0.014 &  0.016$\pm$0.007 &  0.003$\pm$0.005 &                     & (6.05$\pm$1.24)e-14 & (6.20$\pm$1.28)e-14 \\ 
 HD\,300584 & B1Ve & 0\_0\_0		       &2295.3$\pm$98.6  & 4.527$\pm$0.037 & 0.48  &  0.050$\pm$0.009 &                 &  0.003$\pm$0.003 &  0.038$\pm$0.008 &  0.028$\pm$0.008 &                     & (7.52$\pm$1.46)e-14 & (2.25$\pm$0.36)e-13 \\
 QY\,Car & B5Vne & 1\_0\_0	               & 551.2$\pm$14.3  & 3.682$\pm$0.022 & 0.033 &  0.016$\pm$0.005 & 0.002$\pm$0.002 &  0.008$\pm$0.004 &  0.004$\pm$0.003 &  0.003$\pm$0.004 & (5.57$\pm$4.22)e-15 & (1.06$\pm$0.41)e-14 & (1.09$\pm$0.44)e-14 \\ 
 J\,Vel & B5II & 1\_0\_1		       & 352.1$\pm$34.7  & 3.724$\pm$0.086 & 0.027 &  0.039$\pm$0.008 &                 &  0.008$\pm$0.004 &  0.004$\pm$0.003 &  0.002$\pm$0.003 &                     & (1.02$\pm$0.46)e-14 & (1.04$\pm$0.48)e-14 \\ 
 HD\,90563 (\gc) & B2Ve & 1\_0\_0	       &2405.0$\pm$58.0  & 4.326$\pm$0.021 & 0.39  &  0.030$\pm$0.007 & 0.001$\pm$0.002 &  0.001$\pm$0.002 &  0.022$\pm$0.006 &  0.028$\pm$0.008 & (6.67$\pm$9.51)e-14 & (3.95$\pm$1.09)e-14 & (2.12$\pm$0.43)e-13 \\
 HD\,302798 & B3V(e) & 0\_0\_0	               &2357.9$\pm$95.3  & 3.952$\pm$0.035 & 0.32  &  0.037$\pm$0.008 &                 &  0.007$\pm$0.004 &  0.026$\pm$0.006 &  0.009$\pm$0.005 &                     & (5.24$\pm$1.16)e-14 & (7.51$\pm$1.55)e-14 \\
 HD\,310080 & B5e & 0\_0\_0		       &1847.4$\pm$37.1  & 3.319$\pm$0.017 & 0.12  &  0.051$\pm$0.007 & 0.002$\pm$0.002 &  0.019$\pm$0.004 &  0.028$\pm$0.005 &  0.008$\pm$0.004 & (1.63$\pm$1.31)e-14 & (4.73$\pm$0.68)e-14 & (8.99$\pm$1.26)e-14 \\
 HD\,91597 & B7.5IVe & 1\_0\_0		       &3976.2$\pm$225.1 & 3.742$\pm$0.049 & 0.13  &  0.024$\pm$0.006 &                 &  0.017$\pm$0.005 &  0.007$\pm$0.003 &  0.000$\pm$0.001 &                     & (2.20$\pm$0.55)e-14 & (2.24$\pm$0.57)e-14 \\
 V402\,Car & B1e & 2\_2\_0	               &2380.1$\pm$90.1  & 4.100$\pm$0.033 & 0.20  &  0.101$\pm$0.019 &                 &  0.000$\pm$0.000 &  0.000$\pm$0.000 &  0.000$\pm$0.000 &                     & (1.20$\pm$0.22)e-13 & (1.26$\pm$0.23)e-13 \\
 HD\,305483 & B2 & 0\_0\_0	               &2265.2$\pm$58.3  & 3.862$\pm$0.022 & 0.16  &  0.038$\pm$0.008 & 0.018$\pm$0.005 &  0.017$\pm$0.005 &  0.000$\pm$0.002 &  0.002$\pm$0.003 & (1.49$\pm$0.41)e-13 & (1.13$\pm$0.39)e-14 & (1.14$\pm$0.38)e-14 \\
 HD\,93190 & O9.5e & 2\_0\_0		       &2390.8$\pm$183.1 & 5.024$\pm$0.067 & 0.35  &  0.018$\pm$0.006 &                 &  0.004$\pm$0.004 &  0.012$\pm$0.005 &  0.002$\pm$0.003 &                     & (2.53$\pm$0.90)e-14 & (2.59$\pm$0.91)e-14 \\
 HD\,305560 & O9I & 0\_0\_0		       &5598.0$\pm$396.9 & 5.304$\pm$0.062 & 0.36  &  0.065$\pm$0.009 &                 &  0.012$\pm$0.004 &  0.047$\pm$0.008 &  0.027$\pm$0.007 &                     & (9.33$\pm$1.37)e-14 & (2.46$\pm$0.32)e-13 \\
 HD\,93563 & B5III & 0\_0\_0		       & 177.2$\pm$3.4   & 2.806$\pm$0.017 & 0.010 &  0.040$\pm$0.008 & 0.012$\pm$0.004 &  0.023$\pm$0.006 &  0.002$\pm$0.002 &  0.000$\pm$0.001 & (1.80$\pm$0.61)e-14 & (1.83$\pm$0.47)e-14 & (1.87$\pm$0.49)e-14 \\ 
 HD\,93843 & O5III(fc) & 0\_0\_0	       &2320.5$\pm$138.0 & 5.574$\pm$0.052 & 0.21  &  0.189$\pm$0.015 & 0.008$\pm$0.004 &  0.128$\pm$0.013 &  0.041$\pm$0.007 &  0.004$\pm$0.004 & (9.31$\pm$4.38)e-14 & (2.03$\pm$0.18)e-13 & (2.08$\pm$0.18)e-13 \\
 HD\,305627 & B1IIIe & 1\_0\_0		       &4177.5$\pm$287.1 & 4.583$\pm$0.060 & 0.24  &  0.017$\pm$0.005 &                 &  0.009$\pm$0.004 &  0.007$\pm$0.003 &  0.003$\pm$0.003 &                     & (2.02$\pm$0.62)e-14 & (2.07$\pm$0.62)e-14 \\
 HD\,94963 & O7II(f) & 0\_0\_0		       &2715.7$\pm$206.0 & 4.475$\pm$0.066 & 0.11  &  0.153$\pm$0.014 & 0.007$\pm$0.003 &  0.117$\pm$0.012 &  0.025$\pm$0.006 &  0.006$\pm$0.004 & (4.76$\pm$2.03)e-14 & (1.30$\pm$0.12)e-13 & (1.69$\pm$0.16)e-13 \\
 HD\,308104 & B2Ve & 1\_0\_0		       &4351.7$\pm$249.7 & 4.076$\pm$0.050 & 0.25  &  0.023$\pm$0.006 & 0.002$\pm$0.002 &  0.010$\pm$0.005 &  0.010$\pm$0.004 &  0.001$\pm$0.002 & (3.36$\pm$2.93)e-14 & (2.53$\pm$0.77)e-14 & (2.59$\pm$0.81)e-14 \\
 Cl\,Pismis\,17\,3 & B2V & 0\_1\_0             &2634.1$\pm$103.0 & 3.529$\pm$0.034 & 0.27  &  0.217$\pm$0.022 & 0.004$\pm$0.004 &  0.055$\pm$0.012 &  0.137$\pm$0.017 &  0.046$\pm$0.014 & (7.86$\pm$8.10)e-14 & (2.75$\pm$0.30)e-13 & (4.91$\pm$0.52)e-13 \\
 HD\,303763 & Be & 0\_0\_0	               &2295.4$\pm$67.9  & 3.794$\pm$0.026 & 0.22  &  0.046$\pm$0.008 &                 &  0.015$\pm$0.004 &  0.023$\pm$0.005 &  0.017$\pm$0.005 &                     & (5.21$\pm$0.92)e-14 & (1.22$\pm$0.19)e-13 \\
 V353\,Car & B2Ve & 0\_0\_0		       & 913.9$\pm$26.9  & 3.694$\pm$0.026 & 0.075 &  0.037$\pm$0.007 & 0.006$\pm$0.003 &  0.009$\pm$0.004 &  0.020$\pm$0.005 &  0.002$\pm$0.002 & (2.19$\pm$1.03)e-14 & (2.77$\pm$0.62)e-14 & (2.84$\pm$0.64)e-14 \\ 
 HD\,97253 & O5III(f) & 0\_0\_0		       &2302.2$\pm$125.9 & 5.779$\pm$0.047 & 0.27  &  0.276$\pm$0.019 & 0.006$\pm$0.003 &  0.157$\pm$0.015 &  0.077$\pm$0.011 &  0.006$\pm$0.003 & (1.41$\pm$0.71)e-13 & (3.42$\pm$0.27)e-13 & (3.64$\pm$0.28)e-13 \\
 HD\,97535 & B7Vn & 1\_0\_0		       & 239.6$\pm$1.1   & 2.458$\pm$0.004 & 0.10  &  0.093$\pm$0.011 & 0.007$\pm$0.003 &  0.050$\pm$0.008 &  0.033$\pm$0.007 &  0.002$\pm$0.002 & (4.38$\pm$1.88)e-14 & (8.07$\pm$1.02)e-14 & (8.27$\pm$1.04)e-14 \\ 
 KP\,Mus & B2IIIne & 0\_0\_0		       &1415.1$\pm$35.6  & 3.507$\pm$0.022 & 0.16  &  0.099$\pm$0.011 &                 &  0.035$\pm$0.006 &  0.058$\pm$0.008 &  0.023$\pm$0.006 &                     & (1.03$\pm$0.12)e-13 & (2.39$\pm$0.25)e-13 \\
 V870\,Cen & B3Vne: & 1\_0\_0		       &2341.0$\pm$77.4  & 3.542$\pm$0.029 & 0.24  &  0.065$\pm$0.012 & 0.002$\pm$0.002 &  0.036$\pm$0.009 &  0.025$\pm$0.008 &  0.006$\pm$0.006 & (2.45$\pm$3.18)e-14 & (8.14$\pm$1.59)e-14 & (1.10$\pm$0.22)e-13 \\
 HD\,308829 & B8 & 0\_2\_0	               &2540.4$\pm$105.1 & 3.553$\pm$0.036 & 0.37  &  0.150$\pm$0.016 &                 &  0.049$\pm$0.009 &  0.088$\pm$0.012 &  0.014$\pm$0.007 &                     & (2.16$\pm$0.23)e-13 & (2.56$\pm$0.28)e-13 \\
 HD\,102154 & B5IIIe & 1\_0\_0		       &1547.8$\pm$38.9  & 3.196$\pm$0.022 & 0.16  &  0.019$\pm$0.005 &                 &  0.007$\pm$0.003 &  0.010$\pm$0.004 &  0.002$\pm$0.002 &                     & (1.91$\pm$0.54)e-14 & (1.95$\pm$0.55)e-14 \\
 DG\,Cru & B2Vne & 1\_0\_0	               & 898.4$\pm$22.3  & 3.785$\pm$0.022 & 0.098 &  0.021$\pm$0.006 & 0.002$\pm$0.002 &  0.012$\pm$0.004 &  0.007$\pm$0.003 &  0.000$\pm$0.002 & (1.16$\pm$0.94)e-14 & (1.61$\pm$0.46)e-14 & (1.64$\pm$0.51)e-14 \\ 
 V863\,Cen & B5V & 0\_0\_1	               & 134.4$\pm$11.0  & 2.916$\pm$0.071 & 0.0073&  0.239$\pm$0.019 & 0.034$\pm$0.006 &  0.139$\pm$0.015 &  0.043$\pm$0.008 &  0.003$\pm$0.003 & (4.88$\pm$0.85)e-14 & (1.38$\pm$0.13)e-13 & (1.41$\pm$0.13)e-13 \\ 
 $\delta$\,Cen & B2Vne & 0\_0\_2	       & 127.2$\pm$7.6   & 3.923$\pm$0.052 & 0.0067&  2.192$\pm$0.056 &                 &  0.081$\pm$0.007 &  0.017$\pm$0.005 &  0.000$\pm$0.001 &                     & (7.24$\pm$0.67)e-14 & (7.42$\pm$0.70)e-14 \\ 
 V817\,Cen & B3IVe & 2\_0\_0		       & 547.2$\pm$27.3  & 3.873$\pm$0.043 & 0.042 &  0.011$\pm$0.005 &                 &  0.008$\pm$0.004 &  0.002$\pm$0.002 &  0.000$\pm$0.002 &                     & (8.28$\pm$3.97)e-15 & (8.28$\pm$4.37)e-15 \\ 
 HD\,107302 & B8.5IVe & 2\_0\_0		       & 983.0$\pm$14.5  & 2.711$\pm$0.013 & 0.14  &  0.010$\pm$0.005 & 0.001$\pm$0.002 &  0.004$\pm$0.003 &  0.005$\pm$0.003 &  0.003$\pm$0.003 & (8.97$\pm$13.3)e-15 & (8.86$\pm$4.25)e-15 & (8.86$\pm$4.07)e-15 \\ 
 $\zeta$\,Crv & B8V & 1\_0\_0		       & 117.6$\pm$1.2   & 2.267$\pm$0.009 & 0.011 &  0.021$\pm$0.008 & 0.012$\pm$0.005 &  0.005$\pm$0.004 &  0.000$\pm$0.003 &  0.006$\pm$0.005 & (1.88$\pm$0.83)e-14 & (2.43$\pm$2.53)e-15 & (2.46$\pm$1.72)e-15 \\ 
\hline
  \end{tabular}
\end{table}
\end{landscape}
 
\setcounter{table}{0}
\begin{landscape}
\begin{table}
  \scriptsize
  \caption{Continued. } 
  \begin{tabular}{lccccccccccccc}
    \hline
 Name & SpT & flags & $d$ (pc) & $\log(L_{\rm BOL}$ & $N^{\rm ISM}_{\rm H}$ & \multicolumn{5}{c}{Rate (counts\,s$^{-1}$)} & \multicolumn{3}{c}{$F^{ISM\,cor}_{\rm X}$ (erg\,cm$^{-2}$\,s$^{-1}$)} \\  
      &     &       &          &  $/L_{\odot})$           & ($10^{22}$ & (0.2--2.3) & (0.2--0.5) & (0.5--1.) & (1.--2.) & (2.--5.) & (0.2--0.5) & (0.5--2.) & (0.5--5.) \\
      &     &       &          &                               & cm$^{-2}$)  \\      
\hline
 HD\,107609 & B8.5IVe & 2\_0\_0		       & 933.5$\pm$15.0  & 2.611$\pm$0.014 & 0.13  &  0.010$\pm$0.004 & 0.001$\pm$0.001 &  0.006$\pm$0.003 &  0.002$\pm$0.002 &  0.000$\pm$0.002 & (5.58$\pm$9.04)e-15 & (8.95$\pm$4.02)e-15 & (8.95$\pm$4.44)e-15 \\ 
 BZ\,Cru (\gc) & B0.5IVpe & 0\_0\_0	       & 443.5$\pm$16.9  & 4.610$\pm$0.033 & 0.20  &  1.769$\pm$0.048 & 0.031$\pm$0.006 &  0.467$\pm$0.024 &  1.136$\pm$0.038 &  0.387$\pm$0.026 & (3.48$\pm$0.71)e-13 & (1.95$\pm$0.06)e-12 & (3.87$\pm$0.10)e-12 \\ 
 $\mu^{02}$\,Cru & B5Vne & 0\_0\_0              & 120.1$\pm$1.9   & 2.532$\pm$0.014 & 0.010 &  0.319$\pm$0.022 & 0.062$\pm$0.010 &  0.162$\pm$0.016 &  0.087$\pm$0.012 &  0.009$\pm$0.005 & (9.48$\pm$1.49)e-14 & (1.95$\pm$0.15)e-13 & (2.55$\pm$0.20)e-13 \\ 
 HD\,112147 & Be & 0\_0\_0	               &1931.9$\pm$44.5  & 4.059$\pm$0.020 & 0.21  &  0.055$\pm$0.010 &                 &  0.000$\pm$0.001 &  0.052$\pm$0.009 &  0.030$\pm$0.008 &                     & (6.31$\pm$1.09)e-14 & (2.38$\pm$0.36)e-13 \\
 HD\,113573 & B4Ve & 2\_0\_0		       &1163.3$\pm$30.4  & 3.390$\pm$0.023 & 0.13  &  0.018$\pm$0.006 & 0.001$\pm$0.002 &  0.012$\pm$0.005 &  0.005$\pm$0.003 &  0.000$\pm$0.002 & (1.01$\pm$1.54)e-14 & (1.82$\pm$0.62)e-14 & (1.82$\pm$0.64)e-14 \\
 V955\,Cen & B2IVne & 0\_0\_0		       &1083.1$\pm$22.9  & 3.538$\pm$0.018 & 0.23  &  0.256$\pm$0.020 & 0.011$\pm$0.004 &  0.078$\pm$0.011 &  0.154$\pm$0.016 &  0.051$\pm$0.011 & (1.44$\pm$0.56)e-13 & (2.90$\pm$0.24)e-13 & (5.94$\pm$0.46)e-13 \\
 HD\,115805 & B1Vnne & 1\_0\_0		       &2034.8$\pm$67.9  & 4.065$\pm$0.029 & 0.23  &  0.022$\pm$0.007 &                 &  0.009$\pm$0.004 &  0.010$\pm$0.004 &  0.000$\pm$0.002 &                     & (2.47$\pm$0.82)e-14 & (2.54$\pm$0.86)e-14 \\
 V967\,Cen & B0IIe & 1\_0\_0		       &2103.7$\pm$85.4  & 5.224$\pm$0.035 & 0.29  &  0.023$\pm$0.006 & 0.002$\pm$0.002 &  0.007$\pm$0.003 &  0.013$\pm$0.005 &  0.001$\pm$0.002 & (7.79$\pm$6.32)e-14 & (2.94$\pm$0.85)e-14 & (3.01$\pm$0.89)e-14 \\
 LV\,Mus & B2Vne & 0\_0\_0		       &1181.8$\pm$23.7  & 4.052$\pm$0.017 & 0.14  &  0.180$\pm$0.017 & 0.004$\pm$0.003 &  0.058$\pm$0.009 &  0.108$\pm$0.013 &  0.038$\pm$0.009 & (3.40$\pm$2.31)e-14 & (1.79$\pm$0.17)e-13 & (3.83$\pm$0.34)e-13 \\
 HD\,119682 (\gc) & B0Ve & 0\_0\_0	       &1582.4$\pm$65.3  & 4.703$\pm$0.036 & 0.18  &  0.404$\pm$0.026 & 0.021$\pm$0.006 &  0.100$\pm$0.013 &  0.241$\pm$0.023 &  0.071$\pm$0.013 & (2.03$\pm$0.58)e-13 & (3.98$\pm$0.31)e-13 & (7.51$\pm$0.54)e-13 \\
 $\mu$\,Cen & B2Vnpe & 0\_0\_2		       & 155.0$\pm$3.8   & 3.733$\pm$0.022 & 0.0080&  0.532$\pm$0.032 &                 &  0.161$\pm$0.016 &  0.094$\pm$0.014 &  0.015$\pm$0.007 &                     & (1.95$\pm$0.16)e-13 & (2.70$\pm$0.22)e-13 \\ 
 HD\,122669 & B0Ve & 0\_0\_0		       &2112.6$\pm$72.8  & 4.657$\pm$0.030 & 0.25  &  0.097$\pm$0.014 & 0.001$\pm$0.002 &  0.029$\pm$0.008 &  0.059$\pm$0.010 &  0.016$\pm$0.007 & (1.07$\pm$2.55)e-14 & (1.13$\pm$0.16)e-13 & (1.54$\pm$0.21)e-13 \\
 HD\,122691 & Be & 1\_0\_0		       &1952.2$\pm$49.5  & 4.074$\pm$0.022 & 0.23  &  0.013$\pm$0.006 & 0.003$\pm$0.002 &  0.006$\pm$0.004 &  0.004$\pm$0.003 &  0.000$\pm$0.002 & (3.99$\pm$3.21)e-14 & (1.28$\pm$0.63)e-14 & (1.28$\pm$0.71)e-14 \\
 V716\,Cen & B5V & 1\_0\_0		       & 251.7$\pm$3.4   & 3.023$\pm$0.012 & 0.11  &  0.019$\pm$0.007 &                 &  0.003$\pm$0.003 &  0.018$\pm$0.006 &  0.001$\pm$0.003 &                     & (2.18$\pm$0.75)e-14 & (2.23$\pm$0.81)e-14 \\ 
 $\epsilon$\,Aps & B3V & 0\_0\_0	       & 197.6$\pm$8.6   & 3.130$\pm$0.038 & 0.023 &  0.060$\pm$0.009 & 0.013$\pm$0.004 &  0.032$\pm$0.006 &  0.014$\pm$0.005 &  0.001$\pm$0.002 & (2.56$\pm$0.79)e-14 & (3.73$\pm$0.64)e-14 & (3.82$\pm$0.65)e-14 \\ 
 HD\,126986 & B9IVne & 0\_0\_0		       & 606.3$\pm$11.6  & 1.991$\pm$0.017 & 0.10  &  0.028$\pm$0.007 &                 &  0.015$\pm$0.005 &  0.011$\pm$0.005 &  0.000$\pm$0.002 &                     & (2.58$\pm$0.69)e-14 & (2.64$\pm$0.75)e-14 \\ 
 HL\,Lib & B9IVe & 0\_0\_0		       & 313.2$\pm$3.0   & 2.340$\pm$0.008 & 0.034 &  0.058$\pm$0.012 & 0.010$\pm$0.005 &  0.036$\pm$0.009 &  0.009$\pm$0.005 &  0.009$\pm$0.006 & (2.41$\pm$1.20)e-14 & (3.56$\pm$0.82)e-14 & (5.66$\pm$1.29)e-14 \\ 
 CK\,Cir & B2Vne & 0\_0\_0		       & 670.4$\pm$21.3  & 3.752$\pm$0.028 & 0.081 &  0.238$\pm$0.018 & 0.013$\pm$0.004 &  0.081$\pm$0.011 &  0.127$\pm$0.013 &  0.029$\pm$0.008 & (5.50$\pm$1.81)e-14 & (2.02$\pm$0.17)e-13 & (3.34$\pm$0.26)e-13 \\ 
 CQ\,Cir (\gc) & B1Ve & 0\_0\_0		       &1575.2$\pm$46.4  & 3.814$\pm$0.026 & 0.25  &  0.548$\pm$0.030 & 0.011$\pm$0.004 &  0.067$\pm$0.010 &  0.400$\pm$0.026 &  0.222$\pm$0.021 & (1.93$\pm$0.75)e-13 & (6.41$\pm$0.39)e-13 & (1.78$\pm$0.09)e-12 \\
 $\theta$\,Cir & B2III & 1\_0\_0	       & 463.0$\pm$62.2  & 4.107$\pm$0.117 & 0.077 &  0.026$\pm$0.007 & 0.005$\pm$0.003 &  0.012$\pm$0.005 &  0.009$\pm$0.004 &  0.004$\pm$0.004 & (1.79$\pm$1.09)e-14 & (1.72$\pm$0.52)e-14 & (4.20$\pm$1.28)e-14 \\ 
 HD\,135160 & B0.5Ve & 0\_0\_0		       & 515.5$\pm$101.0 & 4.311$\pm$0.170 & 0.090 &  0.212$\pm$0.019 & 0.027$\pm$0.007 &  0.152$\pm$0.016 &  0.022$\pm$0.007 &  0.002$\pm$0.004 & (1.34$\pm$0.32)e-13 & (1.52$\pm$0.15)e-13 & (1.56$\pm$0.16)e-13 \\ 
 HD\,139431 & B2Vne & 1\_0\_0		       & 658.0$\pm$15.2  & 3.576$\pm$0.020 & 0.083 &  0.038$\pm$0.010 & 0.001$\pm$0.002 &  0.010$\pm$0.006 &  0.026$\pm$0.008 &  0.013$\pm$0.008 & (5.58$\pm$9.28)e-15 & (3.35$\pm$0.90)e-14 & (1.11$\pm$0.28)e-13 \\ 
 V1040\,Sco & B2V & 0\_0\_0		       & 142.1$\pm$1.4   & 2.989$\pm$0.009 & 0.067 &  0.189$\pm$0.021 & 0.013$\pm$0.006 &  0.084$\pm$0.014 &  0.086$\pm$0.014 &  0.018$\pm$0.008 & (4.44$\pm$1.90)e-14 & (1.52$\pm$0.18)e-13 & (2.40$\pm$0.27)e-13 \\ 
 HD\,141926 & B2IIIn & 0\_0\_0		       &1317.7$\pm$31.1  & 3.672$\pm$0.021 & 0.21  &  0.100$\pm$0.014 & 0.004$\pm$0.003 &  0.018$\pm$0.006 &  0.072$\pm$0.012 &  0.041$\pm$0.010 & (4.53$\pm$3.81)e-14 & (1.10$\pm$0.16)e-13 & (3.42$\pm$0.43)e-13 \\
 $\delta$\,Sco & B0.3IV & 0\_0\_2              & 150.6$\pm$20.2  & 4.631$\pm$0.116 & 0.086 &  7.823$\pm$0.126 & 0.123$\pm$0.032 &  0.822$\pm$0.038 &  0.137$\pm$0.018 &  0.008$\pm$0.006 & (5.67$\pm$1.48)e-13 & (7.83$\pm$0.34)e-13 & (8.18$\pm$0.36)e-13 \\ 
 HD\,143545 & B1.5Vne & 1\_0\_0		       & 833.5$\pm$12.4  & 3.564$\pm$0.013 & 0.27  &  0.037$\pm$0.009 &                 &  0.002$\pm$0.002 &  0.023$\pm$0.007 &  0.039$\pm$0.010 &                     & (4.07$\pm$1.22)e-14 & (2.28$\pm$0.46)e-13 \\ 
 MQ\,TrA & B1Ve & 0\_0\_0	               & 856.1$\pm$33.2  & 4.078$\pm$0.034 & 0.10  &  0.331$\pm$0.025 & 0.012$\pm$0.005 &  0.119$\pm$0.015 &  0.173$\pm$0.019 &  0.066$\pm$0.013 & (7.53$\pm$3.15)e-14 & (2.93$\pm$0.24)e-13 & (6.34$\pm$0.49)e-13 \\ 
 HD\,143700 & B1.5Vne & 1\_0\_0		       & 970.7$\pm$20.3  & 3.640$\pm$0.018 & 0.27  &  0.032$\pm$0.009 &                 &  0.005$\pm$0.004 &  0.028$\pm$0.008 &  0.006$\pm$0.005 &                     & (4.47$\pm$1.19)e-14 & (4.58$\pm$1.20)e-14 \\ 
 HD\,143343 & B8Vnp & 0\_0\_0		       &1001.0$\pm$17.5  & 2.247$\pm$0.015 & 0.045 &  0.086$\pm$0.013 & 0.004$\pm$0.003 &  0.048$\pm$0.009 &  0.031$\pm$0.008 &  0.002$\pm$0.003 & (1.10$\pm$0.78)e-14 & (6.81$\pm$1.05)e-14 & (6.98$\pm$1.09)e-14 \\
 HD\,144965 & B3Vne & 0\_0\_0		       & 257.4$\pm$1.9   & 2.863$\pm$0.007 & 0.19  &  0.057$\pm$0.012 & 0.002$\pm$0.003 &  0.031$\pm$0.009 &  0.023$\pm$0.008 &  0.000$\pm$0.005 & (2.43$\pm$3.26)e-14 & (5.91$\pm$1.29)e-14 & (6.05$\pm$1.42)e-14 \\ 
 HD\,146596 & B5IVe & 2\_0\_0		       & 830.3$\pm$19.4  & 3.248$\pm$0.020 & 0.084 &  0.018$\pm$0.008 &                 &  0.014$\pm$0.007 &  0.005$\pm$0.004 &  0.000$\pm$0.002 &                     & (1.77$\pm$0.73)e-14 & (1.77$\pm$0.76)e-14 \\ 
 HD\,147302 & B2IIIne & 2\_0\_0		       & 689.9$\pm$13.9  & 3.474$\pm$0.017 & 0.11  &  0.017$\pm$0.007 &                 &  0.011$\pm$0.006 &  0.005$\pm$0.004 &  0.000$\pm$0.002 &                     & (1.64$\pm$0.70)e-14 & (1.64$\pm$0.71)e-14 \\ 
 HD\,148877 & B6Ve & 1\_0\_0		       & 616.5$\pm$9.3   & 2.690$\pm$0.013 & 0.084 &  0.070$\pm$0.014 & 0.006$\pm$0.004 &  0.014$\pm$0.007 &  0.048$\pm$0.011 &  0.000$\pm$0.001 & (2.51$\pm$1.68)e-14 & (7.62$\pm$1.61)e-14 & (7.85$\pm$1.67)e-14 \\ 
 V1063\,Sco & B4IIIe & 1\_0\_0		       & 375.0$\pm$3.9   & 2.830$\pm$0.009 & 0.20  &  0.031$\pm$0.010 &                 &  0.015$\pm$0.007 &  0.017$\pm$0.007 &  0.002$\pm$0.004 &                     & (3.93$\pm$1.18)e-14 & (4.02$\pm$1.24)e-14 \\ 
 HD\,153222 & B1IIe & 0\_0\_0		       &1098.1$\pm$19.6  & 3.814$\pm$0.016 & 0.22  &  0.137$\pm$0.019 & 0.004$\pm$0.003 &  0.036$\pm$0.010 &  0.086$\pm$0.015 &  0.017$\pm$0.008 & (4.35$\pm$4.16)e-14 & (1.52$\pm$0.22)e-13 & (1.95$\pm$0.27)e-13 \\
 HD\,153879 & B1.5Vne & 0\_0\_0		       &1072.9$\pm$30.0  & 3.652$\pm$0.024 & 0.16  &  0.065$\pm$0.014 & 0.002$\pm$0.002 &  0.013$\pm$0.007 &  0.044$\pm$0.010 &  0.009$\pm$0.006 & (1.24$\pm$1.88)e-14 & (6.73$\pm$1.44)e-14 & (9.11$\pm$1.89)e-14 \\
 HD\,154154 & B1Vnne & 0\_0\_0		       &1056.0$\pm$22.3  & 3.949$\pm$0.018 & 0.21  &  0.048$\pm$0.011 & 0.006$\pm$0.004 &  0.015$\pm$0.006 &  0.025$\pm$0.008 &  0.021$\pm$0.009 & (6.84$\pm$4.58)e-14 & (4.95$\pm$1.26)e-14 & (1.82$\pm$0.41)e-13 \\
 HD\,154538 & B3V & 1\_0\_0		       & 663.5$\pm$9.2   & 3.051$\pm$0.012 & 0.17  &  0.030$\pm$0.010 &                 &  0.015$\pm$0.007 &  0.015$\pm$0.006 &  0.014$\pm$0.008 &                     & (3.19$\pm$1.00)e-14 & (1.29$\pm$0.37)e-13 \\ 
 V1075\,Sco & O7.5V((f))z(e) & 0\_0\_0	       &1036.9$\pm$86.6  & 5.382$\pm$0.073 & 0.17  &  0.725$\pm$0.041 & 0.083$\pm$0.014 &  0.536$\pm$0.037 &  0.080$\pm$0.014 &  0.003$\pm$0.003 & (7.19$\pm$1.18)e-13 & (6.40$\pm$0.42)e-13 & (6.54$\pm$0.42)e-13 \\
 HD\,156172 & O9Vnn(e) & 1\_0\_0	       &1177.3$\pm$35.6  & 4.604$\pm$0.026 & 0.31  &  0.025$\pm$0.009 & 0.004$\pm$0.003 &  0.011$\pm$0.006 &  0.008$\pm$0.006 &  0.000$\pm$0.003 & (2.06$\pm$1.57)e-13 & (2.93$\pm$1.27)e-14 & (3.00$\pm$1.37)e-14 \\
 V1077\,Sco & B6Vne & 0\_0\_0		       & 375.0$\pm$4.6   & 3.203$\pm$0.011 & 0.13  &  0.104$\pm$0.017 & 0.010$\pm$0.005 &  0.034$\pm$0.010 &  0.052$\pm$0.012 &  0.016$\pm$0.008 & (7.06$\pm$3.68)e-14 & (9.20$\pm$1.66)e-14 & (1.71$\pm$0.29)e-13 \\ 
 $\gamma$\,Ara & B1Ib & 0\_0\_2		       & 341.3$\pm$18.6  & 4.646$\pm$0.047 & 0.061 &  0.970$\pm$0.046 &                 &  0.416$\pm$0.030 &  0.164$\pm$0.019 &  0.013$\pm$0.007 &                     & (5.08$\pm$0.31)e-13 & (5.50$\pm$0.33)e-13 \\ 
 V750\,Ara (\gc) & B2Vne & 0\_0\_0	       & 972.0$\pm$47.6  & 4.182$\pm$0.043 & 0.081 &  0.475$\pm$0.034 & 0.007$\pm$0.004 &  0.096$\pm$0.016 &  0.344$\pm$0.029 &  0.135$\pm$0.021 & (2.93$\pm$1.85)e-14 & (4.31$\pm$0.32)e-13 & (1.16$\pm$0.08)e-12 \\ 
 V862\,Ara & B7IIIe & 1\_0\_0		       & 427.0$\pm$10.3  & 3.267$\pm$0.021 & 0.055 &  0.026$\pm$0.009 &                 &  0.012$\pm$0.006 &  0.016$\pm$0.006 &  0.003$\pm$0.003 &                     & (2.28$\pm$0.76)e-14 & (2.33$\pm$0.75)e-14 \\ 
 V864\,Ara & B7Vnnpe & 0\_0\_0		       & 233.2$\pm$1.5   & 2.310$\pm$0.005 & 0.035 &  0.620$\pm$0.037 & 0.082$\pm$0.014 &  0.315$\pm$0.026 &  0.196$\pm$0.021 &  0.048$\pm$0.013 & (1.92$\pm$0.32)e-13 & (4.23$\pm$0.28)e-13 & (6.50$\pm$0.42)e-13 \\ 
 V830\,Ara & B2IIpe & 0\_0\_0		       &1558.0$\pm$101.5 & 3.994$\pm$0.057 & 0.10  &  0.129$\pm$0.019 &                 &  0.041$\pm$0.011 &  0.082$\pm$0.015 &  0.014$\pm$0.008 &                     & (1.19$\pm$0.18)e-13 & (1.53$\pm$0.22)e-13 \\
 V1083\,Sco & B2Vne & 0\_0\_0		       &1255.8$\pm$55.9  & 4.126$\pm$0.039 & 0.38  &  0.122$\pm$0.019 & 0.006$\pm$0.004 &  0.011$\pm$0.006 &  0.104$\pm$0.017 &  0.010$\pm$0.007 & (3.32$\pm$2.28)e-13 & (1.71$\pm$0.27)e-13 & (2.43$\pm$0.37)e-13 \\
 HD\,316341 & B0Ve & 0\_0\_0		       &1502.3$\pm$68.3  & 4.317$\pm$0.040 & 0.41  &  0.081$\pm$0.016 &                 &  0.023$\pm$0.008 &  0.044$\pm$0.012 &  0.030$\pm$0.011 &                     & (1.21$\pm$0.27)e-13 & (2.63$\pm$0.50)e-13 \\
 HD\,161774 & B8Vnne & 0\_0\_0		       & 580.8$\pm$6.7   & 2.417$\pm$0.010 & 0.095 &  0.101$\pm$0.017 & 0.008$\pm$0.005 &  0.045$\pm$0.012 &  0.040$\pm$0.011 &  0.010$\pm$0.007 & (4.39$\pm$2.57)e-14 & (8.21$\pm$1.56)e-14 & (1.08$\pm$0.20)e-13 \\ 
 HD\,161807 & O9.7IIInn & 1\_0\_0              &1258.0$\pm$62.5  & 4.646$\pm$0.043 & 0.12  &  0.059$\pm$0.020 & 0.013$\pm$0.006 &  0.039$\pm$0.013 &  0.010$\pm$0.006 &  0.002$\pm$0.006 & (8.69$\pm$4.13)e-14 & (4.42$\pm$1.28)e-14 & (4.51$\pm$1.35)e-14 \\
 $\lambda$\,Pav & B2Ve & 0\_0\_1	       & 338.5$\pm$21.3  & 4.152$\pm$0.055 & 0.034 &  0.348$\pm$0.030 & 0.026$\pm$0.005 &  0.144$\pm$0.019 &  0.128$\pm$0.018 &  0.039$\pm$0.012 & (6.05$\pm$1.27)e-14 & (2.30$\pm$0.22)e-13 & (4.27$\pm$0.40)e-13 \\ 
 $\beta$\,Scl & B9.5IIIpHgMnSi & 0\_0\_1       &  55.9$\pm$0.7   & 1.800$\pm$0.011 & 0.0006&  0.719$\pm$0.048 & 0.138$\pm$0.019 &  0.371$\pm$0.035 &  0.173$\pm$0.024 &  0.012$\pm$0.009 & (1.68$\pm$0.24)e-13 & (4.06$\pm$0.32)e-13 & (4.42$\pm$0.35)e-13 \\ 
\hline
  \end{tabular}
\end{table}
\end{landscape}

\setcounter{table}{0}
\begin{landscape}
\begin{table}
  \scriptsize
  \caption{Continued - this second part provides the X-ray luminosities, X-ray to bolometric ratios, hardness ratios, softness ratios (see text). }
  \begin{tabular}{lccccclccccccc}
    \hline
 Name & SpT & flags & $L_{\rm X}$(0.5--5.) & $\log(L_{\rm X}$ & $HR$ & $SR$ &  Name & SpT & flags & $L_{\rm X}$(0.5--5.) & $\log(L_{\rm X}$ & $HR$ & $SR$\\  
      &     &       & (erg\,s$^{-1}$) & $/L_{\rm BOL})$       &      &      &       &     &       & (erg\,s$^{-1}$) & $/L_{\rm BOL})$ & & \\
\hline
 HD\,19818 & B9.5Vne & 0\_0\_0		       & (3.43$\pm$0.18)e+30 & --4.44$\pm$0.02 & 0.07$\pm$0.07 & 0.28$\pm$0.04 & CD--28\,5235 & B1V & 1\_0\_0		       & (3.87$\pm$1.31)e+31 & --6.20$\pm$0.14 & 0.02$\pm$0.45 &              \\ 
 228\,Eri & B2Vne & 2\_0\_0		       & (8.29$\pm$11.2)e+28 & --8.62$\pm$0.59 & 0.00$\pm$1.61 & 5.02$\pm$7.30 & CD--31\,5475 & B3Vep & 2\_0\_0		       & (4.71$\pm$1.80)e+30 & --6.05$\pm$0.17 & 0.00$\pm$0.55 & 0.46$\pm$0.57\\ 
 $\lambda$\,Eri & B2III(e)p & 0\_0\_1	       & (8.72$\pm$1.75)e+29 & --7.42$\pm$0.08 & 0.57$\pm$0.11 &               & WRAY\,15-162 & Bpe & 2\_0\_0		       & (2.03$\pm$1.01)e+32 & --5.47$\pm$0.21 & 0.82$\pm$0.13 &              \\ 
 AN\,Col & B2Vnpe & 0\_0\_0		       & (3.23$\pm$0.86)e+29 & --7.49$\pm$0.11 & 0.03$\pm$0.36 & 0.42$\pm$0.24 & V420\,Pup & B3Ve & 0\_0\_0		       & (1.83$\pm$0.15)e+31 & --5.69$\pm$0.03 & 0.35$\pm$0.07 & 0.20$\pm$0.05\\ 
 V1369\,Ori & B7Ib/II & 1\_0\_0		       & (3.06$\pm$1.35)e+29 & --7.07$\pm$0.19 & 0.02$\pm$0.60 & 0.29$\pm$0.37 & Cl*\,Ruprecht\,55\,DE\,32 & B0IIIe & 1\_0\_0  & (6.32$\pm$2.78)e+32 & --5.78$\pm$0.12 & 0.63$\pm$0.16 &              \\ 
 120\,Tau & B2IVe & 1\_0\_0		       & (1.69$\pm$1.05)e+29 & --8.13$\pm$0.27 & 0.03$\pm$0.83 & 6.82$\pm$5.58 & HD\,69026 & B1.5Ve & 1\_0\_0		       & (2.33$\pm$0.85)e+30 & --6.91$\pm$0.16 & 0.44$\pm$0.30 & 0.31$\pm$0.53\\ 
 43\,Ori & O9.5IVp & 0\_2\_0		       & (1.98$\pm$0.79)e+31 & --6.25$\pm$0.04 & 0.12$\pm$0.12 & 0.14$\pm$0.04 & HD\,69425 & B1Vpe & 1\_0\_0		       & (1.75$\pm$0.68)e+31 & --6.32$\pm$0.17 & 0.59$\pm$0.22 &              \\ 
 HD\,245310 & B2IIInnpe & 0\_0\_0	       & (1.03$\pm$0.18)e+32 & --5.62$\pm$0.07 & 0.50$\pm$0.12 &               & HD\,71072 & B4IIIe & 0\_0\_0		       & (1.24$\pm$0.42)e+30 & --6.78$\pm$0.15 & 0.02$\pm$0.46 &              \\
 $\zeta$\,Tau (\gc) & B1IVe shell & 0\_0\_2    & (3.98$\pm$0.98)e+30 & --7.01$\pm$0.04 & 0.98$\pm$0.01 &               & HD\,71510 & B3IV & 1\_0\_0		       & (5.55$\pm$2.55)e+28 & --8.05$\pm$0.20 & 0.02$\pm$0.59 & 2.67$\pm$1.53\\ 
 $\omega$\,Ori & B3Ve & 1\_0\_1		       & (3.66$\pm$1.83)e+29 & --8.26$\pm$0.20 & 0.02$\pm$0.63 &               & HD\,72063 & B2Vne & 0\_0\_0		       & (1.89$\pm$0.46)e+31 & --6.00$\pm$0.10 & 0.02$\pm$0.33 &              \\ 
 HD\,39018 & B9 & 1\_0\_0		       & (1.70$\pm$0.74)e+29 & --6.43$\pm$0.19 & 0.02$\pm$0.60 & 0.34$\pm$0.46 & V471\,Car & B5ne & 0\_0\_0		       & (9.58$\pm$1.06)e+30 & --5.86$\pm$0.04 & 0.27$\pm$0.10 & 0.24$\pm$0.08\\ 
 V1167\,Tau & B1Vnne & 0\_0\_0		       & (5.78$\pm$0.91)e+31 & --5.87$\pm$0.06 & 0.58$\pm$0.08 & 0.08$\pm$0.08 & HD\,74401 & B1IIIne & 0\_0\_0		       & (1.11$\pm$0.16)e+32 & --5.52$\pm$0.06 & 0.62$\pm$0.08 & 0.30$\pm$0.21\\
 HD\,250980 & B2ne & 1\_0\_0		       & (1.59$\pm$0.57)e+31 & --6.28$\pm$0.15 & 0.03$\pm$0.49 &               & HD\,74559 & B9Ve & 1\_0\_0		       & (2.45$\pm$0.56)e+30 & --5.41$\pm$0.10 & 0.02$\pm$0.31 &              \\
 HD\,251726 & B1Ve & 1\_0\_0		       & (5.01$\pm$1.57)e+31 & --6.46$\pm$0.13 & 0.35$\pm$0.29 &               & HD\,75081 & B9V & 0\_0\_0	               & (8.06$\pm$1.25)e+29 & --5.97$\pm$0.07 & 0.49$\pm$0.11 & 0.18$\pm$0.07\\
 HD\,42054 (\gc\,cand) & B5Ve & 0\_0\_0	       & (1.38$\pm$0.49)e+31 & --5.48$\pm$0.02 & 0.37$\pm$0.03 & 0.14$\pm$0.02 & HD\,75661 & B2Vne & 2\_0\_0		       & (4.09$\pm$1.82)e+30 & --6.64$\pm$0.19 & 0.00$\pm$0.62 &              \\ 
 HD\,253659 & B0.5Vnne & 1\_0\_0	       & (3.18$\pm$1.08)e+31 & --6.92$\pm$0.14 & 0.32$\pm$0.33 & 1.57$\pm$2.91 & HD\,75925 & B4Vnne & 2\_0\_0		       & (4.08$\pm$1.24)e+30 & --6.15$\pm$0.13 & 0.00$\pm$0.41 &              \\
 HD\,43285 & B6Ve & 0\_0\_0		       & (7.05$\pm$0.97)e+29 & --6.33$\pm$0.06 & 0.02$\pm$0.19 & 0.19$\pm$0.09 & CD--45\,4826 & O9 & 1\_0\_0		       & (8.87$\pm$3.50)e+30 & --6.70$\pm$0.17 & 0.03$\pm$0.58 & 1.16$\pm$1.73\\ 
 HD\,43544 & B2.5Ve & 2\_0\_0		       & (1.17$\pm$0.59)e+29 & --7.77$\pm$0.22 & 0.00$\pm$0.70 &               & E\,Car & B3III & 0\_0\_1	               & (1.13$\pm$0.19)e+30 & --7.47$\pm$0.06 & 0.54$\pm$0.10 &              \\ 
 HD\,254647 & Bpe & 1\_0\_0		       & (1.64$\pm$0.38)e+32 & --5.56$\pm$0.10 & 0.64$\pm$0.12 & 0.34$\pm$0.67 & HD\,79066 & A9IV & 0\_0\_0		       & (3.25$\pm$0.52)e+28 & --5.83$\pm$0.07 & 0.02$\pm$0.22 & 0.63$\pm$0.19\\
 HD\,44506 & B3V & 1\_0\_0	               & (2.31$\pm$1.14)e+29 & --8.04$\pm$0.21 & 0.02$\pm$0.66 & 1.86$\pm$1.38 & HD\,79778 & B2Vne & 0\_0\_0		       & (8.41$\pm$1.00)e+31 & --5.34$\pm$0.04 & 0.54$\pm$0.07 & 0.19$\pm$0.10\\ 
 FR\,CMa (\gc) & B1.5IVe & 0\_0\_0             & (6.00$\pm$0.46)e+31 & --6.02$\pm$0.02 & 0.48$\pm$0.04 & 0.10$\pm$0.02 & HD\,80156 & B8.5IVe & 0\_0\_0		       & (1.27$\pm$0.29)e+30 & --5.56$\pm$0.10 & 0.02$\pm$0.32 &              \\ 
 BD--21$^{\circ}$1449 & Be & 1\_0\_0	       & (7.09$\pm$2.32)e+31 & --5.36$\pm$0.13 & 0.03$\pm$0.39 & 0.90$\pm$0.68 & QQ\,Vel & B2:nep & 1\_0\_0		       & (1.49$\pm$0.43)e+32 & --5.83$\pm$0.12 & 0.42$\pm$0.23 & 0.22$\pm$0.41\\
 PZ\,Gem (\gc) & O9:npe & 0\_0\_0	       & (7.89$\pm$1.46)e+30 & --7.21$\pm$0.08 & 0.02$\pm$0.25 & 0.07$\pm$0.19 & QR\,Vel & B2Vne & 0\_0\_0	               & (3.67$\pm$0.88)e+31 & --5.65$\pm$0.10 & 0.54$\pm$0.15 &              \\ 
 HD\,45995 (\gc) & B1.5Vne & 0\_0\_0	       & (2.98$\pm$0.34)e+31 & --6.28$\pm$0.04 & 0.47$\pm$0.07 & 0.11$\pm$0.04 & HD\,83060 & B2Vnne & 0\_0\_0		       & (3.24$\pm$0.53)e+31 & --5.63$\pm$0.07 & 0.66$\pm$0.08 &              \\ 
 QQ\,Gem & Be & 2\_0\_0			       & (6.62$\pm$4.52)e+29 & --6.85$\pm$0.30 & 0.05$\pm$0.92 &               & I\,Hya & B5V & 0\_0\_1			       & (1.15$\pm$0.13)e+30 & --6.52$\pm$0.04 & 0.11$\pm$0.11 & 0.35$\pm$0.07\\ 
 HD\,46484 & B0.5IVe & 2\_0\_0		       & (3.56$\pm$2.03)e+30 & --7.68$\pm$0.24 & 0.00$\pm$0.77 &               & HD\,84567 & B0IV & 0\_0\_0		       & (4.20$\pm$1.35)e+30 & --7.62$\pm$0.13 & 0.02$\pm$0.40 & 0.70$\pm$0.47\\
 HD\,47054 & B8IVe & 0\_0\_0		       & (2.57$\pm$0.63)e+29 & --6.90$\pm$0.10 & 0.02$\pm$0.33 & 0.72$\pm$0.34 & HD\,85860 & B4Vne & 1\_0\_0		       & (5.39$\pm$2.16)e+29 & --6.86$\pm$0.17 & 0.03$\pm$0.54 & 0.33$\pm$0.50\\ 
 V733\,Mon & B3(II)e & 0\_0\_0		       & (8.02$\pm$1.90)e+31 & --5.81$\pm$0.09 & 0.51$\pm$0.15 & 1.41$\pm$0.75 & HD\,86689 & B0/2ne & 0\_0\_0		       & (9.17$\pm$1.78)e+31 & --5.67$\pm$0.07 & 0.41$\pm$0.13 & 0.32$\pm$0.24\\
 15\,Mon & O7V+B1.5/2V & 0\_0\_1	       & (1.97$\pm$0.57)e+31 & --6.59$\pm$0.02 & 0.06$\pm$0.07 & 0.55$\pm$0.05 & OY Hya & B5Ve & 1\_0\_0		       & (6.89$\pm$1.48)e+29 & --6.68$\pm$0.09 & 0.02$\pm$0.28 &              \\ 
 10\,CMa & B2V & 1\_0\_0		       & (1.86$\pm$1.73)e+29 & --8.57$\pm$0.40 & 0.02$\pm$1.22 & 4.47$\pm$4.57 & HD\,300584 & B1Ve & 0\_0\_0		       & (1.42$\pm$0.26)e+32 & --5.96$\pm$0.07 & 0.67$\pm$0.08 &              \\ 
 HP\,CMa & B1.5Vne & 1\_0\_0		       & (4.57$\pm$2.69)e+29 & --8.13$\pm$0.25 & 0.03$\pm$0.71 & 1.16$\pm$1.03 & QY\,Car & B5Vne & 1\_0\_0	               & (3.95$\pm$1.61)e+29 & --7.67$\pm$0.18 & 0.03$\pm$0.55 & 0.51$\pm$0.44\\ 
 V715\,Mon & B2.5III & 1\_0\_0		       & (1.03$\pm$0.48)e+30 & --7.48$\pm$0.20 & 0.03$\pm$0.63 & 0.72$\pm$0.72 & J\,Vel & B5II & 1\_0\_1		       & (1.55$\pm$0.77)e+29 & --8.12$\pm$0.20 & 0.03$\pm$0.62 &              \\ 
 KS\,CMa & B9e & 0\_0\_0		       & (5.39$\pm$0.70)e+30 & --5.58$\pm$0.05 & 0.02$\pm$0.17 & 0.26$\pm$0.11 & HD\,90563 (\gc) & B2Ve & 1\_0\_0              & (1.47$\pm$0.30)e+32 & --5.74$\pm$0.09 & 0.81$\pm$0.06 & 0.31$\pm$0.45\\ 
 $\kappa$\,CMa & B1.5Ve & 0\_0\_2	       & (1.46$\pm$0.33)e+29 & --8.35$\pm$0.10 & 0.02$\pm$0.30 & 4.71$\pm$1.14 & HD\,302798 & B3V(e) & 0\_0\_0		       & (5.00$\pm$1.11)e+31 & --5.84$\pm$0.09 & 0.30$\pm$0.21 &              \\ 
 HZ\,CMa & B6IV)e+A & 0\_0\_0		       & (1.53$\pm$0.35)e+29 & --7.03$\pm$0.10 & 0.03$\pm$0.31 & 0.84$\pm$0.32 & HD\,310080 & B5e & 0\_0\_0		       & (3.67$\pm$0.53)e+31 & --5.34$\pm$0.06 & 0.47$\pm$0.11 & 0.18$\pm$0.15\\ 
 V742\,Mon & B2IIIe & 0\_0\_0		       & (9.35$\pm$0.85)e+31 & --5.69$\pm$0.03 & 0.39$\pm$0.07 & 0.09$\pm$0.05 & HD\,91597 & B7.5IVe & 1\_0\_0		       & (4.25$\pm$1.17)e+31 & --5.70$\pm$0.11 & 0.02$\pm$0.35 &              \\ 
 V746\,Mon & B1V:nn & 1\_0\_0		       & (1.28$\pm$0.44)e+31 & --7.06$\pm$0.14 & 0.02$\pm$0.43 &               & V402\,Car & B1e & 2\_2\_0	               & (8.51$\pm$1.69)e+31 & --5.75$\pm$0.08 & 0.05$\pm$0.25 &              \\
 HD\,266894 & Be & 1\_0\_0	               & (4.38$\pm$1.94)e+31 & --5.90$\pm$0.17 & 0.03$\pm$0.57 &               & HD\,305483 & B2 & 0\_0\_0	               & (6.97$\pm$2.36)e+30 & --6.60$\pm$0.15 & 0.01$\pm$0.48 & 13.1$\pm$5.7 \\
 HD\,51452 & B0IVe & 2\_0\_0		       & (9.91$\pm$5.07)e+30 & --7.53$\pm$0.21 & 0.00$\pm$0.68 &               & HD\,93190 & O9.5e & 2\_0\_0		       & (1.77$\pm$0.68)e+31 & --7.36$\pm$0.15 & 0.03$\pm$0.49 &              \\
 19\,Mon & B2Vn(e) & 0\_0\_0		       & (7.03$\pm$1.84)e+29 & --7.62$\pm$0.10 & 0.02$\pm$0.33 & 1.33$\pm$0.51 & HD\,305560 & O9I & 0\_0\_0		       & (9.22$\pm$1.78)e+32 & --5.92$\pm$0.06 & 0.62$\pm$0.07 &              \\ 
 27\,CMa & B4Ve sh & 0\_0\_1		       & (1.81$\pm$0.65)e+31 & --6.44$\pm$0.05 & 0.81$\pm$0.04 & 0.07$\pm$0.02 & HD\,93563 & B5III & 0\_0\_0		       & (7.03$\pm$1.88)e+28 & --7.54$\pm$0.11 & 0.02$\pm$0.36 & 0.96$\pm$0.41\\ 
 HD\,55806 & B7IIIe & 1\_0\_0		       & (2.57$\pm$0.90)e+30 & --5.75$\pm$0.15 & 0.03$\pm$0.48 & 0.55$\pm$0.46 & HD\,93843 & O5III(fc) & 0\_0\_0	       & (1.34$\pm$0.20)e+32 & --7.03$\pm$0.04 & 0.02$\pm$0.12 & 0.45$\pm$0.21\\ 
 NV\,Pup & B2V+B3IVne & 1\_0\_1		       & (1.52$\pm$0.49)e+29 & --8.04$\pm$0.13 & 0.03$\pm$0.43 &               & HD\,305627 & B1IIIe & 1\_0\_0		       & (4.32$\pm$1.42)e+31 & --6.53$\pm$0.13 & 0.02$\pm$0.42 &              \\ 
 HD\,57682 & O9.2IV & 0\_0\_0		       & (6.32$\pm$1.21)e+31 & --6.51$\pm$0.03 & 0.02$\pm$0.11 & 0.24$\pm$0.07 & HD\,94963 & O7II(f) & 0\_0\_0		       & (1.49$\pm$0.27)e+32 & --5.89$\pm$0.04 & 0.23$\pm$0.10 & 0.28$\pm$0.12\\
 HD\,57910 & B5V & 1\_0\_0	               & (1.56$\pm$0.76)e+30 & --6.12$\pm$0.21 & 0.03$\pm$0.65 & 0.49$\pm$0.65 & HD\,308104 & B2Ve & 1\_0\_0		       & (5.87$\pm$1.95)e+31 & --5.89$\pm$0.14 & 0.02$\pm$0.42 & 1.30$\pm$1.20\\
 NO\,CMa & B3V & 1\_0\_0		       & (2.00$\pm$0.96)e+29 & --7.83$\pm$0.21 & 0.02$\pm$0.63 & 1.27$\pm$0.89 & Cl\,Pismis\,17\,3 & B2V & 0\_1\_0             & (4.08$\pm$0.54)e+32 & --4.50$\pm$0.05 & 0.44$\pm$0.08 & 0.16$\pm$0.17\\ 
 FY\,CMa & B1IIe & 0\_0\_0	               & (2.03$\pm$0.43)e+30 & --7.44$\pm$0.09 & 0.02$\pm$0.27 & 0.69$\pm$0.28 & HD\,303763 & Be & 0\_0\_0	               & (7.67$\pm$1.28)e+31 & --5.49$\pm$0.07 & 0.57$\pm$0.10 &              \\ 
 RY\,Gem & A0V:e+K0II & 0\_0\_0		       & (1.12$\pm$0.26)e+30 & --5.30$\pm$0.10 & 0.02$\pm$0.32 & 0.36$\pm$0.22 & V353\,Car & B2Ve & 0\_0\_0		       & (2.84$\pm$0.67)e+30 & --6.82$\pm$0.10 & 0.02$\pm$0.31 & 0.77$\pm$0.40\\ 
 V378\,Pup & B2Ve & 2\_0\_0		       & (3.59$\pm$1.81)e+29 & --7.91$\pm$0.22 & 0.03$\pm$0.70 & 0.61$\pm$0.90 & HD\,97253 & O5III(f) & 0\_0\_0		       & (2.31$\pm$0.31)e+32 & --7.00$\pm$0.03 & 0.06$\pm$0.10 & 0.39$\pm$0.20\\ 
 BN\,Gem & O8Vpev & 0\_0\_0		       & (1.48$\pm$0.51)e+31 & --7.51$\pm$0.13 & 0.03$\pm$0.42 & 1.74$\pm$0.96 & HD\,97535 & B7Vn & 1\_0\_0		       & (5.68$\pm$0.72)e+29 & --6.29$\pm$0.05 & 0.02$\pm$0.17 & 0.53$\pm$0.24\\
 HD\,62367 & B8V & 1\_0\_0	               & (4.34$\pm$1.74)e+29 & --6.58$\pm$0.17 & 0.02$\pm$0.54 &               & KP\,Mus & B2IIIne & 0\_0\_0		       & (5.74$\pm$0.67)e+31 & --5.33$\pm$0.05 & 0.57$\pm$0.07 &              \\ 
 HD\,62780 & O9/B0e & 1\_0\_0		       & (1.58$\pm$0.52)e+32 & --6.38$\pm$0.14 & 0.72$\pm$0.13 & 0.43$\pm$0.93 & V870\,Cen & B3Vne: & 1\_0\_0		       & (7.22$\pm$1.51)e+31 & --5.27$\pm$0.09 & 0.26$\pm$0.21 & 0.22$\pm$0.29\\
 CD--23\,6121 & Be & 0\_0\_0		       & (2.45$\pm$0.67)e+32 & --5.58$\pm$0.10 & 0.32$\pm$0.23 &               & HD\,308829 & B8 & 0\_2\_0	               & (1.98$\pm$0.27)e+32 & --4.84$\pm$0.05 & 0.16$\pm$0.13 &              \\
 V392\,Pup & B5V & 0\_0\_0	               & (1.16$\pm$0.29)e+29 & --7.13$\pm$0.10 & 0.02$\pm$0.32 & 0.13$\pm$0.13 & HD\,102154 & B5IIIe & 1\_0\_0		       & (5.60$\pm$1.61)e+30 & --6.03$\pm$0.12 & 0.02$\pm$0.39 &              \\ 
 $o$\,Pup & B1IVe & 0\_0\_1   	               & (6.61$\pm$2.42)e+29 & --8.20$\pm$0.13 & 0.02$\pm$0.41 & 1.49$\pm$0.60 & DG\,Cru & B2Vne & 1\_0\_0	               & (1.59$\pm$0.50)e+30 & --7.17$\pm$0.14 & 0.02$\pm$0.41 & 0.71$\pm$0.61\\ 
 CD--29\,5159 & B0Ve & 0\_0\_0		       & (1.29$\pm$0.16)e+33 & --5.48$\pm$0.04 & 0.63$\pm$0.05 &               & V863\,Cen & B5V & 0\_0\_1	               & (3.06$\pm$0.58)e+29 & --7.01$\pm$0.04 & 0.02$\pm$0.13 & 0.35$\pm$0.07\\
 V374\,Car & B3Vn & 0\_0\_0		       & (1.67$\pm$0.19)e+30 & --6.88$\pm$0.05 & 0.18$\pm$0.12 & 0.26$\pm$0.09 & $\delta$\,Cen & B2Vne & 0\_0\_2	       & (1.44$\pm$0.22)e+29 & --8.35$\pm$0.04 & 0.02$\pm$0.13 &              \\ 
 V408\,Pup & B2.5IIIe & 1\_0\_0		       & (2.58$\pm$0.91)e+31 & --6.15$\pm$0.14 & 0.02$\pm$0.46 & 0.57$\pm$0.67 & V817\,Cen & B3IVe & 2\_0\_0		       & (2.97$\pm$1.59)e+29 & --7.98$\pm$0.23 & 0.00$\pm$0.71 &              \\
\hline                                                                                                                         
  \end{tabular}                                                                                                                
\end{table}
\end{landscape}

\setcounter{table}{0}
\begin{table*}
  \scriptsize
  \caption{Continued. } 
  \begin{tabular}{lcccccc}
    \hline
 Name & SpT & flags & $L_{\rm X}$(0.5--5.) & $\log(L_{\rm X}$ & $HR$ & $SR$ \\  
      &     &       & (erg\,s$^{-1}$) & $/L_{\rm BOL})$       &      &      \\
\hline
 HD\,107302 & B8.5IVe & 2\_0\_0		       & (1.03$\pm$0.47)e+30 & --6.28$\pm$0.20 & 0.00$\pm$0.66 & 1.01$\pm$1.57\\
 $\zeta$\,Crv & B8V & 1\_0\_0		       & (4.07$\pm$2.85)e+27 & --8.24$\pm$0.30 & 0.01$\pm$1.24 & 7.64$\pm$6.34\\
 HD\,107609 & B8.5IVe & 2\_0\_0		       & (9.34$\pm$4.64)e+29 & --6.22$\pm$0.22 & 0.00$\pm$0.67 & 0.62$\pm$1.06 \\ 
 BZ\,Cru (\gc) & B0.5IVpe & 0\_0\_0	       & (9.11$\pm$0.73)e+31 & --6.23$\pm$0.01 & 0.50$\pm$0.02 & 0.09$\pm$0.02 \\ 
 $\mu^{02}$\,Cru & B5Vne & 0\_0\_0              & (4.40$\pm$0.37)e+29 & --6.47$\pm$0.03 & 0.23$\pm$0.08 & 0.37$\pm$0.07 \\ 
 HD\,112147 & Be & 0\_0\_0	               & (1.06$\pm$0.17)e+32 & --5.62$\pm$0.07 & 0.74$\pm$0.06 &               \\
 HD\,113573 & B4Ve & 2\_0\_0		       & (2.95$\pm$1.05)e+30 & --6.50$\pm$0.15 & 0.00$\pm$0.49 & 0.56$\pm$0.87 \\
 V955\,Cen & B2IVne & 0\_0\_0		       & (8.34$\pm$0.74)e+31 & --5.20$\pm$0.03 & 0.51$\pm$0.06 & 0.24$\pm$0.10 \\
 HD\,115805 & B1Vnne & 1\_0\_0		       & (1.26$\pm$0.44)e+31 & --6.55$\pm$0.15 & 0.03$\pm$0.46 &               \\
 V967\,Cen & B0IIe & 1\_0\_0		       & (1.60$\pm$0.49)e+31 & --7.60$\pm$0.13 & 0.02$\pm$0.40 & 2.59$\pm$2.23 \\
 LV\,Mus & B2Vne & 0\_0\_0		       & (6.41$\pm$0.63)e+31 & --5.83$\pm$0.04 & 0.53$\pm$0.06 & 0.09$\pm$0.06 \\
 HD\,119682 (\gc) & B0Ve & 0\_0\_0	       & (2.25$\pm$0.25)e+32 & --5.93$\pm$0.03 & 0.47$\pm$0.06 & 0.27$\pm$0.08 \\
 $\mu$\,Cen & B2Vnpe & 0\_0\_2		       & (7.75$\pm$0.75)e+29 & --7.43$\pm$0.04 & 0.28$\pm$0.08 &               \\ 
 HD\,122669 & B0Ve & 0\_0\_0		       & (8.25$\pm$1.27)e+31 & --6.32$\pm$0.06 & 0.27$\pm$0.15 & 0.07$\pm$0.17 \\
 HD\,122691 & Be & 1\_0\_0		       & (5.86$\pm$3.23)e+30 & --6.89$\pm$0.24 & 0.00$\pm$0.74 & 3.10$\pm$3.02 \\
 V716\,Cen & B5V & 1\_0\_0		       & (1.69$\pm$0.61)e+29 & --7.38$\pm$0.16 & 0.03$\pm$0.49 &               \\ 
 $\epsilon$\,Aps & B3V & 0\_0\_0	       & (1.78$\pm$0.34)e+29 & --7.46$\pm$0.07 & 0.02$\pm$0.24 & 0.67$\pm$0.24 \\ 
 HD\,126986 & B9IVne & 0\_0\_0		       & (1.16$\pm$0.33)e+30 & --5.51$\pm$0.12 & 0.02$\pm$0.38 &               \\ 
 HL\,Lib & B9IVe & 0\_0\_0		       & (6.64$\pm$1.52)e+29 & --6.10$\pm$0.10 & 0.37$\pm$0.20 & 0.43$\pm$0.23 \\ 
 CK\,Cir & B2Vne & 0\_0\_0		       & (1.80$\pm$0.18)e+31 & --6.08$\pm$0.03 & 0.39$\pm$0.07 & 0.16$\pm$0.06 \\ 
 CQ\,Cir (\gc) & B1Ve & 0\_0\_0		       & (5.28$\pm$0.41)e+32 & --4.67$\pm$0.02 & 0.64$\pm$0.03 & 0.11$\pm$0.04 \\
 $\theta$\,Cir & B2III & 1\_0\_0	       & (1.08$\pm$0.44)e+30 & --7.66$\pm$0.13 & 0.59$\pm$0.18 & 0.43$\pm$0.29 \\ 
 HD\,135160 & B0.5Ve & 0\_0\_0		       & (4.96$\pm$2.01)e+30 & --7.20$\pm$0.04 & 0.02$\pm$0.14 & 0.86$\pm$0.23 \\ 
 HD\,139431 & B2Vne & 1\_0\_0		       & (5.76$\pm$1.47)e+30 & --6.40$\pm$0.11 & 0.70$\pm$0.11 & 0.05$\pm$0.08 \\ 
 V1040\,Sco & B2V & 0\_0\_0		       & (5.81$\pm$0.67)e+29 & --6.81$\pm$0.05 & 0.37$\pm$0.10 & 0.18$\pm$0.08 \\ 
 HD\,141926 & B2IIIn & 0\_0\_0		       & (7.11$\pm$0.96)e+31 & --5.40$\pm$0.05 & 0.68$\pm$0.06 & 0.13$\pm$0.11 \\
 $\delta$\,Sco & B0.3IV & 0\_0\_2              & (2.22$\pm$0.60)e+30 & --7.87$\pm$0.02 & 0.04$\pm$0.06 & 0.69$\pm$0.18 \\ 
 HD\,143545 & B1.5Vne & 1\_0\_0		       & (1.90$\pm$0.38)e+31 & --5.87$\pm$0.09 & 0.82$\pm$0.06 &               \\ 
 MQ\,TrA & B1Ve & 0\_0\_0	               & (5.56$\pm$0.61)e+31 & --5.92$\pm$0.03 & 0.54$\pm$0.05 & 0.12$\pm$0.05 \\ 
 HD\,143700 & B1.5Vne & 1\_0\_0		       & (5.17$\pm$1.37)e+30 & --6.51$\pm$0.11 & 0.02$\pm$0.36 &               \\ 
 HD\,143343 & B8Vnp & 0\_0\_0		       & (8.37$\pm$1.34)e+30 & --4.91$\pm$0.07 & 0.02$\pm$0.21 & 0.16$\pm$0.11 \\
 HD\,144965 & B3Vne & 0\_0\_0		       & (4.79$\pm$1.13)e+29 & --6.77$\pm$0.10 & 0.02$\pm$0.31 & 0.40$\pm$0.55 \\ 
 HD\,146596 & B5IVe & 2\_0\_0		       & (1.46$\pm$0.63)e+30 & --6.67$\pm$0.19 & 0.00$\pm$0.60 &               \\ 
 HD\,147302 & B2IIIne & 2\_0\_0		       & (9.32$\pm$4.07)e+29 & --7.09$\pm$0.19 & 0.00$\pm$0.61 &               \\ 
 HD\,148877 & B6Ve & 1\_0\_0		       & (3.57$\pm$0.77)e+30 & --5.72$\pm$0.09 & 0.03$\pm$0.29 & 0.32$\pm$0.22 \\ 
 V1063\,Sco & B4IIIe & 1\_0\_0		       & (6.77$\pm$2.10)e+29 & --6.58$\pm$0.13 & 0.02$\pm$0.42 &               \\ 
 HD\,153222 & B1IIe & 0\_0\_0		       & (2.81$\pm$0.40)e+31 & --5.95$\pm$0.06 & 0.22$\pm$0.16 & 0.22$\pm$0.22 \\
 HD\,153879 & B1.5Vne & 0\_0\_0		       & (1.26$\pm$0.27)e+31 & --6.14$\pm$0.09 & 0.26$\pm$0.22 & 0.14$\pm$0.21 \\
 HD\,154154 & B1Vnne & 0\_0\_0		       & (2.43$\pm$0.55)e+31 & --6.15$\pm$0.10 & 0.73$\pm$0.09 & 0.38$\pm$0.26 \\
 HD\,154538 & B3V & 1\_0\_0		       & (6.81$\pm$1.96)e+30 & --5.80$\pm$0.12 & 0.75$\pm$0.10 &               \\ 
 V1075\,Sco & O7.5V((f))z(e) & 0\_0\_0	       & (8.42$\pm$1.51)e+31 & --7.04$\pm$0.03 & 0.02$\pm$0.09 & 1.10$\pm$0.19 \\
 HD\,156172 & O9Vnn(e) & 1\_0\_0	       & (4.97$\pm$2.29)e+30 & --7.49$\pm$0.20 & 0.02$\pm$0.61 & 6.87$\pm$6.09 \\
 V1077\,Sco & B6Vne & 0\_0\_0		       & (2.88$\pm$0.50)e+30 & --6.33$\pm$0.07 & 0.46$\pm$0.13 & 0.41$\pm$0.23 \\ 
 $\gamma$\,Ara & B1Ib & 0\_0\_2		       & (7.67$\pm$0.96)e+30 & --7.34$\pm$0.03 & 0.08$\pm$0.08 &               \\ 
 V750\,Ara (\gc) & B2Vne & 0\_0\_0	       & (1.31$\pm$0.16)e+32 & --5.65$\pm$0.03 & 0.63$\pm$0.04 & 0.03$\pm$0.02 \\ 
 V862\,Ara & B7IIIe & 1\_0\_0		       & (5.08$\pm$1.66)e+29 & --7.14$\pm$0.14 & 0.02$\pm$0.45 &               \\ 
 V864\,Ara & B7Vnnpe & 0\_0\_0		       & (4.23$\pm$0.28)e+30 & --5.27$\pm$0.03 & 0.35$\pm$0.06 & 0.30$\pm$0.05 \\ 
 V830\,Ara & B2IIpe & 0\_0\_0		       & (4.44$\pm$0.86)e+31 & --5.93$\pm$0.06 & 0.22$\pm$0.16 &               \\
 V1083\,Sco & B2Vne & 0\_0\_0		       & (4.58$\pm$0.81)e+31 & --6.05$\pm$0.07 & 0.29$\pm$0.15 & 1.37$\pm$0.96 \\
 HD\,316341 & B0Ve & 0\_0\_0		       & (7.11$\pm$1.49)e+31 & --6.05$\pm$0.08 & 0.54$\pm$0.13 &               \\
 HD\,161774 & B8Vnne & 0\_0\_0		       & (4.35$\pm$0.82)e+30 & --5.36$\pm$0.08 & 0.24$\pm$0.20 & 0.41$\pm$0.25 \\ 
 HD\,161807 & O9.7IIInn & 1\_0\_0              & (8.55$\pm$2.69)e+30 & --7.30$\pm$0.13 & 0.02$\pm$0.41 & 1.92$\pm$1.08 \\
 $\lambda$\,Pav & B2Ve & 0\_0\_1	       & (5.85$\pm$0.92)e+30 & --6.97$\pm$0.04 & 0.46$\pm$0.07 & 0.14$\pm$0.03 \\ 
 $\beta$\,Scl & B9.5IIIpHgMnSi & 0\_0\_1       & (1.65$\pm$0.14)e+29 & --6.16$\pm$0.03 & 0.08$\pm$0.10 & 0.38$\pm$0.06 \\ 
\hline
  \end{tabular}

{\scriptsize The known \gc\ stars are identified by parenthesis in the first column. Rates in 0.2--2.3, 1.--2., and 2.--5.\,keV are observed rates while rates in 0.2--0.5 and 0.5--1.\,keV have been corrected for optical loading. Fluxes are fluxes after correction for interstellar absorption. Flag is `q\_m\_o' where q is the detection quality 0 (very good) -- 2 (fair), m informs on the multiplicity/extent (0: clean, 1: potentially confused, 2: detected as mildly extended), and o informs on the optical contamination (0: none/negligible, 1: low/moderate, 2: medium). A missing value means the value could not be determined. $HR$ is the fraction of the total (0.5--5.\,keV) flux being the hard (2.--5.\,keV), calculated by $1-F^{ISM\,cor}_{\rm X}(0.5-2.)/F^{ISM\,cor}_{\rm X}(0.5-5.)$, and $SR$ is the ratio between the flux in 0.2--0.5\,keV and the flux in 0.5--5.\,keV.}
\end{table*}


\bsp	
\label{lastpage}
\end{document}